\documentclass[11pt]{article}
\usepackage[utf8]{inputenc}
\usepackage{amsfonts}
\usepackage{braket}
\usepackage[bottom]{footmisc}
\usepackage{xcolor}
\usepackage{qcircuit}
\usepackage{amsmath}
\usepackage{algorithm}
\usepackage{algpseudocode}
\algrenewcommand\algorithmicrequire{\textbf{Input:}}
\algrenewcommand\algorithmicensure{\textbf{Output:}}
\usepackage{mathabx}
\usepackage{enumerate}
\usepackage{hyperref}
\usepackage{amsthm}
\usepackage{tikz}
\usetikzlibrary{arrows.meta, positioning}
\usetikzlibrary{mindmap}
\usepackage{caption}
\usepackage{cleveref}
\usepackage{authblk}
\usepackage{mdframed}
\usepackage{lipsum}
\usepackage{amsmath}
\usepackage{amssymb}
\usepackage[title]{appendix}
\usetikzlibrary{matrix,decorations.pathreplacing}
\usepackage{mleftright}
\usepackage{amssymb}
\usepackage{algorithm}
\usepackage{algpseudocode}
\algrenewcommand\algorithmicrequire{\textbf{Input:}}
\algrenewcommand\algorithmicensure{\textbf{Output:}}
\usepackage{mathabx}
\usepackage{geometry}
\usepackage{comment}
\usepackage{arydshln}
\usepackage{subcaption}

\newtheorem{proposition}{Proposition}

\newtheorem{definition}{Definition}

\newtheorem*{claim*}{Claim}
\newtheorem{corollary}{Corollary}

\usepackage{mathtools}
\usepackage{stackengine}

\newtheorem{theorem}{Theorem}
\usepackage{tikz-cd}
\usetikzlibrary{cd}
\allowdisplaybreaks
\usepackage{tikz}
\usetikzlibrary{shapes.geometric, arrows}
\tikzstyle{startstop} = [rectangle, rounded corners, minimum width=3cm, minimum height=1cm,text centered, draw=black, fill=red!30]
\tikzstyle{arrow} = [thick,->,>=stealth]
\usepackage{geometry}
\def\be{\begin{eqnarray}}

\allowdisplaybreaks
\makeatletter

\title{\textbf{The Pre-Optimization Structure of QAOA: Polynomial Equivalence of Average and Optimal Probabilities}
}

\author[1,3]{Guilherme Adamatti Bridi\textsuperscript{*}}
\author[1,2]{Franklin de Lima Marquezino\textsuperscript{\dag}}
\author[3]{Andris Ambainis\textsuperscript{\ddag}}

\affil[1]{PESC-COPPE, Universidade Federal do Rio de Janeiro, Brazil}
\affil[2]{Duque de Caxias Campus, Universidade Federal do Rio de Janeiro, Brazil}
\affil[3]{Center for Quantum Computing Science, University of Latvia, Latvia}

\date{\today}

\begin{document}

\maketitle

\begingroup
\renewcommand{\thefootnote}{}
\footnotetext{%
\textsuperscript{*}\texttt{gabridi@cos.ufrj.br}
\quad
\textsuperscript{\dag}\texttt{franklin@cos.ufrj.br}
\quad
\textsuperscript{\ddag}\texttt{andris.ambainis@lu.lv}
}
\endgroup

\begin{abstract}
Two fundamental aspects in understanding the performance of the Quantum Approximate Optimization Algorithm (QAOA) are problem structure and variational parameter optimization. In this work, we develop a theoretical framework connecting these two aspects. We first derive a general bound on how much variational optimization can improve the probability of measuring a configuration in an arbitrary subset of computational basis states, relating the maximum achievable probability to its average over variational parameters. For NPO-PB problems with polynomially many distinct mixer eigenvalues and constant circuit depth, this bound establishes that optimization can improve the average probability by at most a polynomial factor. This raises a fundamental question: to what extent is variational optimization necessary for sampling desirable solutions? We show that, under suitable efficiency assumptions on circuit implementation and parameter sampling, QAOA with randomly chosen parameters can achieve, up to a constant factor, the cumulative success probabilities of variationally trained QAOA, with only a polynomial multiplicative overhead in circuit executions. In particular, under these conditions, efficient sampling of desirable solutions can be achieved without variational training whenever variationally optimized QAOA achieves inverse-polynomial success probability. Beyond its algorithmic implications, the mean--max bound reveals that, under polynomial mean--max proximity, a pre-optimization structural layer determines the asymptotic scaling of optimal measurement probabilities up to polynomial factors. For single-layer QAOA initialized in the uniform superposition state, we explicitly characterize this layer for the measurement probability of a reference state using an exact expression for the average probability that captures the interplay between problem structure and mixer dynamics through the overlap of two matrices: one encoding the problem structure, derived from the joint contributions of the problem and mixer operators, and the other reflecting the intrinsic nature of the mixer. We establish a mixer-dependent capacity bound and identify cancellations between constructive and destructive structural contributions as a mechanism for additional superpolynomial suppression of the average measurement probability relative to the mixer capacity. These structural mechanisms are illustrated through analytically tractable examples involving transverse field and Grover mixers.
\end{abstract}


\section{Introduction}

Understanding the performance of the Quantum Approximate Optimization Algorithm (QAOA) remains a central challenge in quantum optimization. QAOA~\cite{farhi2014quantum} combines alternating evolutions under a problem Hamiltonian, which encodes the objective function, and a mixer Hamiltonian, which enables transitions between computational basis states. Through the interference generated by these evolutions, the algorithm seeks to enhance the probability of measuring good solutions. However, the extent to which this enhancement can be achieved depends on several interconnected factors, including the properties and constraints of the combinatorial optimization problem and its encoding; the choice of mixer; the initial state; the circuit depth; parameter initialization and optimization strategies; and noise and hardware constraints. These factors shape both the potential advantages and the limitations of QAOA, making their interplay central to its theoretical analysis.

In this context, empirical investigations have played a central role in assessing QAOA performance across a wide range of combinatorial optimization problems and algorithmic settings. These studies encompass several often overlapping lines of investigation, with empirical performance evaluations across different problem instances, performance metrics, and circuit depths forming a common component of much of this literature. Such investigations also address more specific aspects of QAOA performance, including comparisons across QAOA variants~\cite{golden2023numerical,golden2023quantum,matwiejew2024quantum,slate2021quantum}, assessments of QAOA's competitiveness against classical optimization methods~\cite{crooks2018performance,lykov2023sampling,shaydulin2024evidence}, and comparisons with alternative quantum optimization approaches, particularly quantum annealing~\cite{willsch2020benchmarking,zhou2020quantum}. Experimental implementations have further explored QAOA performance under noise and hardware constraints~\cite{willsch2020benchmarking,pagano2020quantum,harrigan2021quantum,shaydulin2024evidence}. On the other hand, obtaining general analytical characterizations of QAOA performance remains challenging, given the complexity of its quantum dynamics and variational optimization landscape. Nevertheless, substantial theoretical progress has been made in several directions, including rigorous performance guarantees for specific optimization problems~\cite{farhi2014quantum,wurtz2021maxcut,farhi2025lower} and theoretical limitations on the performance achievable by QAOA~\cite{bravyi2020obstacles,mbeng2019quantum,farhi2020typical,farhi2020worstcase,chou2021limitations,benchasattabuse2025lower,bridi2024analytical,xie2025performance}, as well as analytical characterizations of algorithmic performance~\cite{farhi2022sherrington,basso2022iterative,wang2018quantum,ng2024analytical,headley2023problem,headley_thesis,bridi2024analytical,hadfield2023analytical,diezvalle2024connection}.

These advances provide important insights into QAOA performance, although understanding the interplay between its different ingredients remains an active area of research. In the following, we focus on two complementary aspects: the role of problem structure and the influence of variational parameter optimization on QAOA performance.

\subsection{Problem structure}

In classical optimization, the structure of a problem encompasses properties and regularities of its objective function and search space that can influence the difficulty of finding optimal solutions. In QAOA, however, identifying which aspects of this structure are relevant to algorithmic performance, and how their influence depends on the problem and mixer Hamiltonians, remains an important theoretical question. Existing theoretical studies have provided insights into particular aspects of problem structure that may be relevant to QAOA dynamics and performance, through distinct analytical frameworks. Locality-based analyses have established how finite-depth QAOA depends on restricted neighborhoods of the interaction graph, providing insights into its performance and limitations~\cite{farhi2014quantum,farhi2020typical,farhi2020worstcase,chou2021limitations}. Classical symmetries have been exploited to identify equivalent configurations and invariant subspaces, construct symmetry-preserving mixers, and reduce QAOA representations~\cite{shaydulin2020classical,tsvelikhovskiy2024,tsvelikhovskiy2026}. Related algebraic approaches based on dynamical Lie algebras characterize the structure generated by the problem and mixer Hamiltonians, providing insights into the expressivity and controllability of the ansatz~\cite{larocca2022diagnosing,larocca2023theory,meyer2023exploiting,ragone2024lie,allcock2024dynamical,adamatti2026expressivity,tsvelikhovskiy2025provable}. The choice of mixer provides another important perspective. generalized QAOA frameworks allow mixing operators to be tailored to the constraints of the optimization problem~\cite{hadfield2019quantum}. Another perspective on mixing operators arises from their connection to continuous-time quantum walks~\cite{farhi1998quantum}, which offers a geometric interpretation of amplitude propagation through the search space~\cite{marsh2019quantum,marsh2020combinatorial}. This connection has also motivated analyses of amplitude transfer on mixer graphs~\cite{matwiejew2024quantum}. Complementary analytical approaches have investigated QAOA dynamics through cost gradient operators~\cite{hadfield2023analytical} and exact expressions for expectation values in different QAOA variants~\cite{ng2024analytical}. Other studies have examined the influence of the statistical properties of objective functions~\cite{headley2023problem,bridi2024analytical,headley_thesis} and the connection between quantum interference and thermal-like sampling behavior~\cite{diezvalle2024connection,diezvalle2026universal}.

Taken together, these developments highlight that the structure relevant to QAOA performance depends not only on the optimization problem, but also on how the chosen mixer acts on the search space. As a variational version of Grover's diffusion operator~\cite{grover1996fast}, it has become an important theoretical benchmark for understanding the limitations of amplitude amplification in QAOA. In particular, Bridi and Marquezino~\cite{bridi2024analytical} established Grover-like bounds on amplitude amplification, with an alternative proof subsequently presented by Xie et al.~\cite{xie2025performance}. From a structural perspective, the Grover mixer is equivalent, up to a global phase and a rescaling of the variational parameter, to a continuous-time quantum walk on the complete graph~\cite{marsh2020combinatorial, bennett2021quantum1}. Therefore, its dynamics distinguish only two classes of configurations relative to a target state $z$: the target itself and all remaining configurations, with no finer hierarchy of neighborhoods or distances.

This observation suggests that achieving amplitude amplification beyond Grover-like quadratic scaling may require mixers capable of resolving richer structural distinctions than the binary partition induced by the Grover mixer. In this sense, the Grover mixer serves as a natural baseline for assessing improvements beyond unstructured search, a perspective already recognized by Refs~\cite{golden2023numerical, golden2023quantum}. This raises a more fundamental question: which aspects of problem structure are actually accessible to QAOA dynamics, and how does this accessible structure influence the amplitude amplification process?

\subsection{Variational parameters and optimization}

Beyond the structural properties of the problem, QAOA performance also depends on how its variational parameters are chosen, motivating extensive research into their optimization and the associated challenges. In the standard variational approach, a classical optimizer iteratively adjusts the circuit parameters to optimize an objective function, typically the expectation value of the problem Hamiltonian. Various classical optimization methods have been developed or employed for this purpose, including gradient-based approaches~\cite{zhou2020quantum,wierichs2020avoiding,sweke2020stochastic} and gradient-free methods~\cite{nakanishi2020sequential,parrish2019jacobi}. Complementary strategies have also been developed to facilitate parameter selection and optimization, including informed parameter initialization~\cite{sack2021quantum}, the identification of parameter concentration patterns~\cite{akshay2021parameter}, interpolation across circuit depths and Fourier-based parameterizations~\cite{zhou2020quantum}, and parameter transfer between related problem instances~\cite{galda2023similarity}. Nevertheless, the optimization of variational parameters faces several fundamental challenges. The optimization landscape is generally nonconvex~\cite{zhou2020quantum} and may contain numerous local minima~\cite{rajakumar2024trainability}, posing challenges for practical parameter optimization. More fundamentally, Bittel and Kliesch~\cite{bittel2021training} established that training variational quantum algorithms is NP-hard in the worst case, highlighting the computational complexity of parameter optimization. Another important challenge is the occurrence of barren plateaus, first identified by McClean et al.~\cite{mcclean2018barren}, in which gradients become exponentially suppressed under certain conditions. Subsequent studies have investigated the dependence of this phenomenon on the cost function and circuit architecture~\cite{cerezo2021cost}, as well as its emergence under noise~\cite{wang2021noise}.

These challenges make it difficult to assess the potential improvement offered by variational optimization, while also motivating a closer examination of whether such optimization is essential for achieving good performance. This raises a natural question: how much can optimizing the variational parameters improve the probability of measuring desirable solutions compared with a baseline such as randomly chosen parameters? In particular, is this improvement sufficiently large to make parameter optimization essential for efficiently sampling desirable solutions?

\subsection{Our contributions}

In this work, we investigate the fundamental limits of the benefits of variational parameter optimization and how these limits reveal the role of problem structure in determining QAOA performance. We develop a theoretical framework connecting these two aspects, establishing a link between the potential improvement offered by variational optimization and the structure underlying QAOA performance. 

In Sec.~\ref{sec:structure}, we begin by investigating the notion of problem structure in QAOA, particularly by identifying what constitutes problem structure from the perspective of the algorithm. Focusing on single-layer QAOA initialized in the uniform superposition state, we develop a structural framework to express the amplitude of a reference state, based on the coarsest partition of the search space that preserves the individual contributions of the problem and mixer operators. In particular, when the mixer Hamiltonian is the adjacency matrix of a graph, the mixer contribution admits an interpretation in terms of walks on that graph. This leads to a joint distribution that captures the problem structure visible to the algorithm. This representation also allows us to investigate statistical dependence between the problem and mixer coordinates and establish necessary conditions for exponential amplitude amplification.

In Sec.~\ref{sec:optimization}, we investigate how much variational optimization can improve the probability of measuring configurations in a reference subset by establishing a general bound relating this probability averaged over variational parameters to the maximum probability achievable through optimization. This bound applies to arbitrary QAOA depth, arbitrary initial states, and arbitrary subsets of computational basis configurations. The key consequence of this result is that, for NPO-PB problems with mixer Hamiltonians having polynomially many distinct eigenvalues and constant QAOA depth, the bound establishes polynomial mean--max proximity, i.e., optimization can improve the average probability by at most a polynomial factor. We then explore the algorithmic consequences of polynomial mean--max proximity to assess the extent to which variational optimization is necessary for efficiently sampling desirable solutions. First, we establish an anticoncentration guarantee showing that parameter choices yielding probabilities at least a constant fraction of the average cannot be superpolynomially rare under polynomial mean--max proximity. We then show that a number of independent QAOA executions, each using freshly sampled variational parameters, determined by the mean--max proximity factor and thus polynomial under polynomial mean--max proximity, suffices to guarantee a cumulative success probability bounded below by a constant fraction of the maximum single-execution success probability. Extending this comparison to variationally trained QAOA, we establish that random-parameter sampling can achieve at least a constant fraction of the cumulative success probability of variationally trained QAOA, with only a polynomial overhead in circuit executions. Consequently, under the stated efficiency and parameter-sampling assumptions, whenever a variationally trained QAOA circuit achieves inverse-polynomial success probability, random-parameter QAOA can efficiently sample desirable configurations without requiring variational training.

In Sec.~\ref{sec:pre-optimization}, we bring together the two aspects investigated in the preceding sections: problem structure and variational parameter optimization. Under polynomial mean--max proximity, the average measurement probability determines the asymptotic scaling of the optimal measurement probability up to polynomial factors, revealing the existence of a pre-optimization structural layer. We then return to single-layer QAOA initialized in the uniform superposition state and focus on the measurement probability of a single reference state, under the conditions for polynomial mean--max proximity established for NPO-PB problems and mixers with polynomially many distinct eigenvalues. This allows us to revisit the question of problem structure raised in Sec.~\ref{sec:structure}, now focusing on its role in determining the average measurement probability in this particular regime. We derive an exact expression for the average probability that captures the interplay between problem structure and mixer dynamics through the overlap between a problem structure Gram matrix and a mixer accessibility matrix. Under polynomial mean--max proximity, this Gram matrix provides a reduced representation of the original problem structure that is sufficient to determine the asymptotic scaling of optimal measurement probabilities up to polynomial factors. We then establish an intrinsic mixer-dependent upper bound on the structural overlap, which may or may not be superpolynomially small, and use it as a reference for assessing the role of problem structure. By classifying the contributions to the structural overlap according to their signs, we distinguish constructive, blind, and destructive contributions. We show that, when the numbers of distinct problem costs and mixer eigenvalues are polynomially bounded, the constructive contribution reaches the mixer capacity up to polynomial factors. Consequently, any additional superpolynomial suppression of the structural overlap relative to the mixer capacity must arise from cancellations between constructive and destructive contributions. We illustrate these distinct regimes through analytically tractable examples involving transverse field and Grover mixers, showing how the interplay between problem structure and mixer dynamics leads to different patterns of constructive and destructive contributions to the structural overlap.

\section{Background and notation}
\label{sec:background}

We consider a combinatorial optimization problem defined by an objective function $c:\mathcal{S}\rightarrow\mathbb{R}$, where $n$ denotes the input size of a problem instance and $\mathcal{S}$ is a finite search space of size $N=|\mathcal{S}|$. Without loss of generality, we assume that the objective is to maximize $c$. Depending on the formulation, $\mathcal{S}$ may represent the full configuration space (unconstrained) or a restricted set of feasible solutions (constrained). In particular, $N=2^n$ when $\mathcal{S}=\{0,1\}^n$, but this relation is not assumed in general. Let $\mathcal{H}_{\mathcal{S}}$ denote the Hilbert space spanned by the computational basis states $\{|x\rangle:x\in\mathcal{S}\}$.

The Quantum Approximate Optimization Algorithm (QAOA) operates on the Hilbert space $\mathcal{H}_{\mathcal{S}}$, preparing a parameterized quantum state from an initial state $|\psi_0\rangle$ through alternating applications of a problem operator $U_P(\gamma)$ and a mixing operator $U_M(\beta)$. At depth $p$, the resulting state is
\begin{equation}
    |\psi^{(p)}\rangle = U_M(\beta_p)U_P(\gamma_p)\cdots U_M(\beta_1)U_P(\gamma_1)|\psi_0\rangle,
\end{equation}
where $\boldsymbol{\gamma}=(\gamma_1,\ldots,\gamma_p)$ and $\boldsymbol{\beta}=(\beta_1,\ldots,\beta_p)$ are variational parameters. The operators are defined by $U_P(\gamma)=e^{-i\gamma H_C}$ and $U_M(\beta)=e^{-i\beta H_M}$, where $H_C$ is the problem Hamiltonian, diagonal in the computational basis, satisfying $H_C|x\rangle=c(x)|x\rangle$ and $H_M$ is the mixer Hamiltonian, which generates transitions between computational basis states. 

When $H_M$ is associated with the adjacency matrix of an undirected, unweighted graph whose vertices represent the configurations in $\mathcal{S}$, the mixing operator $U_M(\beta)$ can be interpreted as a continuous-time quantum walk (CTQW) on that graph, with $\beta$ playing the role of evolution time. Two important examples are the transverse field mixer and the Grover mixer, both of which will serve as recurring examples throughout this work. For $\mathcal{S}=\{0,1\}^n$, the transverse field mixer is generated by $H_M=\sum_{i=1}^n X_i$, corresponding to the adjacency matrix of the $n$-dimensional hypercube, where transitions connect configurations differing by a single bit. In contrast, the Grover mixer is generated by $H_M=|s\rangle\langle s|$, with $|s\rangle=N^{-1/2}\sum_{x\in\mathcal{S}}|x\rangle$, and induces global mixing among all configurations in the search space. Up to a global phase and a rescaling of the evolution parameter, the Grover mixer is equivalent to a CTQW generated by the adjacency matrix of the complete graph.

In Sections~\ref{sec:structure} and~\ref{sec:pre-optimization}, we take the initial state $|\psi_0\rangle$ to be the uniform superposition $|s\rangle=\frac{1}{\sqrt{N}}\sum_{x\in\mathcal{S}}|x\rangle$. In Section~\ref{sec:optimization}, we consider arbitrary normalized initial states.

For a reference configuration $z\in\mathcal{S}$, we denote its measurement amplitude by $\psi_z^{(p)}=\langle z|\psi^{(p)}\rangle$ and its corresponding probability by $q_z^{(p)}=|\psi_z^{(p)}|^2$. We define the average measurement probability as
\begin{equation}
\overline{q}_z^{(p)}
=
\lim_{T_\gamma,T_\beta\to\infty}
\frac{1}{T_\gamma^p T_\beta^p}
\int_{[0,T_\gamma]^p}
\int_{[0,T_\beta]^p}
q_z^{(p)}(\boldsymbol{\gamma},\boldsymbol{\beta})
\,d^p\boldsymbol{\beta}\,d^p\boldsymbol{\gamma}.
\end{equation}
The average is taken independently over all variational parameters in the infinite-time limit, without requiring periodicity or integrality of the Hamiltonian spectra. We denote the corresponding averages over the problem and mixer parameters by $\mathbb{E}_\gamma$ and $\mathbb{E}_\beta$, respectively. In particular, for $p=1$, we simplify the notation to $\psi_z$, $q_z$, and $\overline{q}_z$, respectively. Let $A\subseteq\mathcal{S}$ be any subset of computational basis configurations. We denote the total probability of measuring a configuration in $A$ by
\begin{equation}
q_A^{(p)}(\boldsymbol{\gamma},\boldsymbol{\beta})
=
\sum_{z\in A}
q_z^{(p)}(\boldsymbol{\gamma},\boldsymbol{\beta}),
\end{equation}
and its parameter average by $\overline{q}_A^{(p)}=\sum_{z\in A}\overline{q}_z^{(p)}$.

In this work, we explore the class NPO-PB, consisting of NP optimization problems whose objective functions take values in a discrete range polynomially bounded in the input size \cite{adamatti2026expressivity, marsh2019quantum}. In particular, the number of distinct objective values is polynomially bounded in the input size.

\section{What problem structure does QAOA actually see?}
\label{sec:structure}

Problem structure is an essential ingredient of our approach to understanding QAOA performance. To understand how this structure influences the algorithm's performance, we must first identify what structure QAOA actually sees. This is the central question addressed in this section.

\subsection{A minimal perspective on QAOA structure}

Our strategy is to work at the most fundamental level possible, avoiding unnecessary conceptual complexity in order to isolate the basic mechanism of QAOA dynamics. A natural starting point is the amplitude of a single reference state $z$, $\psi_z^{(p)}$. This is a linear object, unlike observables such as the expectation value of the cost Hamiltonian, which introduce additional sums over the solution space. The study of expectation values is standard in QAOA literature, but beginning with them is not necessary for identifying the basic interference mechanism. By starting from a single amplitude, we can keep the contributions of the problem and mixer operators visible, while retaining a direct connection to measurement probabilities and, ultimately, QAOA performance.

Following the same reasoning, we naturally focus on $p=1$. A single QAOA layer is the simplest setting in which the cost-dependent phases and mixer-dependent transitions interact.  The goal is not to claim that the general problem reduces to depth one, but to understand the elementary mechanism first and identify what additional difficulties arise when the depth increases. In this setting, one can write the amplitude as
\begin{equation}
\psi_z =\frac1{\sqrt N}\sum_{x\in\mathcal{S}} B_\beta(z,x)e^{-i\gamma c(x)},
\label{eq:unif_amplitude}
\end{equation}
where $B_\beta(z,x)=\bra{z}U_M(\beta)\ket{x}$. Each summand contains a mixer response $B_\beta(z,x)$ and a problem phase $e^{-i\gamma c(x)}$. We refer to $B_\beta(z,x)$ as the mixer factor and $e^{-i\gamma c(x)}$ as the problem factor.

\subsection{The coarsest structural partition}

The amplitude in Eq.~\eqref{eq:unif_amplitude} can be interpreted as an expectation over uniformly sampled configurations,
\begin{equation}
    \psi_z=\sqrt{N}\,\mathbb{E}_{x\sim\mathrm{Unif}(\mathcal{S})}\left[B_\beta(z,x)e^{-i\gamma c(x)}\right].
\end{equation}
One could argue that problem structure is already encoded in the individual configurations of the search space. While formally correct, this observation provides little insight into how that structure influences QAOA dynamics. The key question is which distinctions between configurations are relevant to the evolution generated by the problem and mixer operators.

To identify these distinctions, we observe that each configuration contributes to the amplitude through the mixer factor $B_\beta(z,x)$ and the problem factor $e^{-i\gamma c(x)}$. Configurations sharing both factors make identical contributions. We therefore seek the coarsest partition of $\mathcal{S}$ that preserves these factors separately for all $\beta$ and $\gamma$, rather than only at particular parameter values. The following theorem establishes the uniqueness of this partition and characterizes its classes for arbitrary problem and Hermitian mixer Hamiltonians.

\begin{theorem}[Coarsest structural partition]
\label{thm:coarsest}
Let $H_M$ be a finite-dimensional Hermitian mixer Hamiltonian acting on the search space $\mathcal{S}$, and let $z$ be a reference configuration. We define $d$ as the number of distinct eigenvalues of $H_M$. The unique coarsest partition of $\mathcal{S}$ preserving the problem and mixer factors individually for all $\gamma$ and $\beta$ is given by
\begin{equation}
    \mathcal{S} = \bigsqcup_{j,k}\mathcal{S}_{j,k}(z), \quad \mathcal{S}_{j,k}(z) = \{x\in\mathcal{S}:J(x)=j,\ K(x)=k\},
\end{equation}
where the coordinates $(J,K)$ are characterized by
\begin{equation}
    J(x)=J(y)\iff c(x)=c(y),
\end{equation}
and
\begin{equation}
    K(x)=K(y)\iff \langle z|H_M^r|x\rangle=\langle z|H_M^r|y\rangle
    \quad\text{for all }r=0,\ldots,d-1.
\end{equation}
We refer to the equivalence classes induced by $J$ and $K$ as problem classes and mixer classes, respectively, and to their intersections $\mathcal{S}_{j,k}(z)$ as structural classes. We denote the numbers of problem and mixer classes by $m=|J|$ and $\ell=|K|$, respectively.
\end{theorem}

\begin{proof}
We first consider the problem factor. Since $e^{-i\gamma c(x)}$ depends on a configuration only through its cost, configurations satisfying $c(x)=c(y)$ have identical problem responses for every $\gamma$. Conversely, if $e^{-i\gamma c(x)}=e^{-i\gamma c(y)}$ for every $\gamma$, then $c(x)=c(y)$.

For the mixer factor, expanding the matrix exponential gives
\begin{equation}
    B_\beta(z,x)=\sum_{r=0}^{\infty}\frac{(-i\beta)^r}{r!}\langle z|H_M^r|x\rangle.
\end{equation}
Suppose that $B_\beta(z,x)=B_\beta(z,y)$ for every $\beta$. Equality of these functions implies equality of their derivatives at $\beta=0$. In particular,
\begin{equation}
    \left.\frac{d^r}{d\beta^r}B_\beta(z,x)\right|_{\beta=0}=(-i)^r\langle z|H_M^r|x\rangle,
\end{equation}
and therefore $\langle z|H_M^r|x\rangle=\langle z|H_M^r|y\rangle$ for every $r\geq 0$, and in particular for $r=0,\ldots,d-1$.

Conversely, suppose that $\langle z|H_M^r|x\rangle=\langle z|H_M^r|y\rangle$ for $r=0,\ldots,d-1$. By the spectral decomposition $H_M=\sum_{j=1}^{d}\lambda_j\Pi_j$, we have
\begin{equation}
    \langle z|H_M^r|x\rangle=\sum_{j=1}^{d}\lambda_j^r\langle z|\Pi_j|x\rangle.
\end{equation}
Hence, equality of these matrix elements implies
\begin{equation}
    \sum_{j=1}^{d}\lambda_j^r\left(\langle z|\Pi_j|x\rangle-\langle z|\Pi_j|y\rangle\right)=0,\qquad r=0,\ldots,d-1.
\end{equation}
Since the associated Vandermonde matrix is invertible, it follows that $\langle z|\Pi_j|x\rangle=\langle z|\Pi_j|y\rangle$ for every $j$. Consequently,
\begin{equation}
    B_\beta(z,x)=\sum_{j=1}^{d}e^{-i\beta\lambda_j}\langle z|\Pi_j|x\rangle=\sum_{j=1}^{d}e^{-i\beta\lambda_j}\langle z|\Pi_j|y\rangle=B_\beta(z,y)
\end{equation}
for every $\beta$. Thus, $K$ characterizes exactly the configurations with identical mixer responses for all $\beta$.

Their joint partition is therefore the unique coarsest partition preserving both factors simultaneously.
\end{proof}

Note that $m$ is the number of distinct cost values, which coincides with the number of distinct eigenvalues of the problem Hamiltonian $H_C$, since it is diagonal in the computational basis. In contrast, the number of mixer classes $\ell$ does not necessarily coincide with the number of distinct eigenvalues $d$ of $H_M$. An important special case of Theorem~\ref{thm:coarsest} arises when the mixer Hamiltonian is the adjacency matrix of an undirected, unweighted graph $G$. In this setting, the matrix element $\langle z|H_M^r|x\rangle$ counts the number of walks of length $r$ connecting $z$ and $x$ in $G$, denoted by $w_r^G(z,x)$. This yields the following combinatorial characterization of the mixer classes.

\begin{corollary}[Walk-count characterization]
\label{cor:walk_counts}
Let $G$ be an undirected, unweighted graph with adjacency Hamiltonian $H_M$, and let $d$ be the number of distinct eigenvalues of $H_M$. Then the mixer classes of Theorem~\ref{thm:coarsest} are characterized by
\begin{equation}
    K(x)=K(y)\iff w_r^G(z,x)=w_r^G(z,y)
    \quad\text{for all }r=0,\ldots,d-1.
\end{equation}
\end{corollary}

It is important to mention that the equivalence between walk-count vectors and spectral projector entries is already known in algebraic graph theory (see Eqs.~(1) and (2) of Dalfó et al.~\cite{dalfo2011almost} and Lemma~2.2 of Fiol~\cite{fiol2016quotient}). In addition, this characterization is also related to the subshell decomposition introduced by Matwiejew and Wang~\cite{matwiejew2024quantum}, which exploits graph automorphisms to identify configurations with identical continuous-time quantum walk amplitudes. In particular, vertices belonging to the same orbit under automorphisms fixing $z$ necessarily have identical walk counts relative to $z$, and hence belong to the same $K$ class. However, this condition is sufficient but not necessary. To see this, consider the graph $G=K_1\vee(C_3\sqcup C_4)$, with $z$ denoting the central vertex. All seven remaining vertices have identical walk counts relative to $z$ at every length, yet vertices in $C_3$ and $C_4$ cannot be related by any automorphism fixing $z$. Thus, distinct automorphism orbits may belong to the same $K$ class, showing that propagation equivalence can be strictly coarser than equivalence under graph automorphisms.

For our two canonical mixers, our characterization coincides exactly with the subshell decomposition based on graph symmetries. For the transverse field mixer, the subshells are the Hamming-distance shells $K(x)=d_H(z,x)$, which are also precisely the classes distinguished by the walk counts $w_r^G(z,x)$. In particular,
\begin{equation} \label{eq:TF_mixerfactors}
B_\beta(z,x)=(-i\sin\beta)^k(\cos\beta)^{n-k}, \qquad k=d_H(z,x).
\end{equation}
The Hamming-distance structure has also been explored in different contexts of QAOA, appearing as a natural framework for analyzing the dynamics of the transverse field mixer~\cite{sud2024parameter,diezvalle2024connection,hadfield2023analytical,headley_thesis}. For the Grover mixer, both constructions yield only two classes: the singleton $\{z\}$ and its complement $\mathcal{S}\setminus\{z\}$. Their mixer factors are
\begin{equation} \label{eq:GM_mixerfactors}
B_\beta(z,x)=
\begin{cases}
1+\dfrac{e^{-i\beta}-1}{N}, & x=z,\\[6pt]
\dfrac{e^{-i\beta}-1}{N}, & x\neq z.
\end{cases}
\end{equation}

\subsection{The joint structural distribution}

The coarsest structural partition established in Theorem~\ref{thm:coarsest} allows us to define a mathematical object that encodes the problem structure visible to QAOA.

\begin{definition}[Joint structural distribution]
At depth one, the problem structure visible to QAOA relative to a reference configuration $z$ is encoded in the joint distribution $P^{(z)}(J,K)$, defined by
\begin{equation}
P^{(z)}(j,k)=\Pr(J=j,K=k)
=\frac{|\mathcal{S}_{j,k}(z)|}{N}.
\end{equation}
\end{definition}

Indeed, the QAOA amplitude can be written exactly in terms of this joint distribution as
\begin{equation}\label{eq:amplitude}
    \psi_z=\sqrt{N}\sum_{j,k}P^{(z)}(j,k)B_\beta^{(z)}(k)e^{-i\gamma j} = \sqrt{N}\,\mathbb{E}_{P^{(z)}(J,K)}\left[B_\beta^{(z)}(K)e^{-i\gamma J}\right],
\end{equation}
where $B_\beta^{(z)}(k)$ denotes the common mixer response associated with the class $K=k$. For the transverse field mixer, the joint distribution $P^{(z)}(J,K)$ reduces to the joint distribution of cost and Hamming distance, an object previously considered in Refs.~\cite{diezvalle2024connection,sud2024parameter}. In particular, Díez-Valle et al.~\cite{diezvalle2024connection} provided a physical interpretation of this distribution as a manifestation of the energy-level structure of the underlying spin system. Our perspective is complementary: the joint distribution arises from the coarsest partition preserving the problem and mixer factors, providing an exact characterization of the structure distinguishable by these factors.

This interpretation follows directly from the structural representation of the amplitude. The relevant information about the objective function is not its complete assignment of costs to individual configurations, but how those costs are distributed across the equivalence classes induced by the mixer. Indeed, $P^{(z)}(J,K)$ retains only the cardinalities $|\mathcal{S}_{j,k}(z)|$, so distinct objective functions inducing the same joint distribution yield identical amplitudes $\psi_z$ for all $\beta$ and $\gamma$, provided the mixer and reference configuration $z$ are fixed. Thus, this representation provides a precise and general notion of problem structure. Note that specific structural features, such as symmetry or locality, may ultimately influence $P^{(z)}(J,K)$.

Finally, this representation provides a clear separation between the classical statistical structure of the instance, encoded by $P^{(z)}(J,K)$, and the quantum dynamical factors induced by the operators, given by $B_\beta^{(z)}(K)e^{-i\gamma J}$. The joint distribution determines the weights assigned to the contributions from each structural class, while the dynamical factors govern how these weighted contributions interfere as the variational parameters change.

\subsection{Independent and correlated contributions}

Having established what problem structure is visible to QAOA, we now begin to investigate its role in amplitude amplification. The expectation representation of the amplitude allows us to separate the contribution associated with statistical independence between the problem and mixer coordinates from that arising from their dependence. Decomposing the joint expectation into the product of marginal expectations and a covariance term, we obtain
\begin{equation}
    \psi_z=\psi_z^{\mathrm{ind}}+\psi_z^{\mathrm{corr}},
    \label{eq:split}
\end{equation}
where
\begin{align}
    \psi_z^{\mathrm{ind}} &= \sqrt{N}\,\mathbb{E}_{P_K^{(z)}}[B_\beta^{(z)}(K)]\,\mathbb{E}_{P_J^{(z)}}[e^{-i\gamma J}],
    \label{eq:ind}\\
    \psi_z^{\mathrm{corr}} &= \sqrt{N}\,\mathrm{Cov}_P^{(z)}\left(B_\beta^{(z)}(K),e^{-i\gamma J}\right).
    \label{eq:corr}
\end{align}
Here, the covariance is understood in the bilinear sense, without complex conjugation. The first term depends only on the marginal distributions, whereas the second captures the contribution arising from statistical dependence between the problem and mixer coordinates.

To understand the independent contribution, consider first its mixer factor. By definition,
\begin{equation}
    \mathbb{E}_{P_K^{(z)}}[B_\beta^{(z)}(K)]=\sum_k P_K^{(z)}(k)B_\beta^{(z)}(k)=\frac{1}{\sqrt{N}}\langle z|U_M(\beta)|s\rangle.
\end{equation}
This quantity characterizes the intrinsic bias of the mixer toward the reference configuration $z$, independently of the objective function. In particular, if $|\langle z|U_M(\beta)|s\rangle|=1/\sqrt{N}$, the mixer preserves the uniform measurement probability $1/N$ at $z$, neither favoring nor suppressing this configuration relative to uniform sampling. Under this condition, it follows that $|\psi_z^{\mathrm{ind}}|\leq 1/\sqrt{N}$, so the independent contribution cannot produce asymptotic amplification beyond the uniform amplitude scale. This motivates the following definition.

\begin{definition}[Unbiased mixers] \label{def:unbiased}
A mixer $U_M(\beta)$ is said to be unbiased with respect to a reference configuration $z$ if
\begin{equation}
    |\langle z|U_M(\beta)|s\rangle|=\frac{1}{\sqrt{N}}, \quad \forall \beta.
\end{equation}
\end{definition}
When $H_M$ is the adjacency matrix of a graph, a sufficient condition for the mixer to be unbiased with respect to every vertex is that the graph be regular, since the uniform superposition is then an eigenstate of $H_M$. Both the transverse field and Grover mixers satisfy this condition. However, unbiasedness is not a necessary property of mixers, which may instead be designed to exhibit intrinsic bias toward particular solutions. 

For unbiased mixers, amplification beyond the uniform amplitude scale therefore requires a nonzero correlation-dependent contribution $\psi_z^{\mathrm{corr}}$, which arises from the covariance between the problem and mixer factors under $P^{(z)}(J,K)$.


\subsection{Statistical dependence as a necessary resource for amplitude amplification}
The decomposition of the QAOA amplitude raises a natural question: to what extent are the problem and mixer coordinates statistically dependent? If $J$ and $K$ were independent, the correlated contribution would vanish identically, regardless of the variational parameters. Statistical dependence is therefore necessary for a nonzero correlated contribution.

One can consider different measures to quantify this dependence, including information-theoretic quantities such as mutual information. However, a particularly useful characterization emerges directly from the correlated contribution itself. Analogously to the joint distribution $P^{(z)}(J,K)$, this contribution can be expressed in terms of a structural weighting function as
\begin{equation}
    \psi_z^{\mathrm{corr}}=\sqrt{N}\sum_{j,k}D^{(z)}(j,k)B_\beta^{(z)}(k)e^{-i\gamma j},
\end{equation}
where
\begin{equation}
    D^{(z)}(j,k)=P^{(z)}(j,k)-P_J^{(z)}(j)P_K^{(z)}(k).
    \label{eq:D}
\end{equation}
The function $D^{(z)}(j,k)$ measures the deviation of the joint distribution from statistical independence, vanishing identically if and only if $J$ and $K$ are independent. It captures how statistical dependence is organized across the structural coordinates $(J,K)$.

The practical significance of the dependence function lies in the fact that $\|D^{(z)}\|_1$ directly bounds the magnitude of the correlated contribution to the QAOA amplitude. In particular, applying the triangle inequality and using $|B_\beta^{(z)}(k)|\leq 1$ and $|e^{-i\gamma j}|=1$, we obtain the following bound.

\begin{proposition}[Dependence bound]
For single-layer QAOA initialized in the uniform superposition $|s\rangle$, the correlation-dependent contribution satisfies, for any reference configuration $z$ and any variational parameters $(\beta,\gamma)$,
\begin{equation}
    |\psi_z^{\mathrm{corr}}|\leq\sqrt{N}\,\|D^{(z)}\|_1,
    \label{eq:dependence_bound}
\end{equation}
where $\|D^{(z)}\|_1=\sum_{j,k}|D^{(z)}(j,k)|$.
\end{proposition}

Beyond its role in bounding amplitude amplification, $\|D^{(z)}\|_1$ provides a natural, parameter-independent measure of how much structure is available for QAOA to exploit. Observe that a similar argument applied directly to the full amplitude would yield only a trivial bound, since $\|P^{(z)}\|_1=1$. In contrast, $\|D^{(z)}\|_1$ can, in principle, vanish for statistically independent distributions. However, the joint distribution considered here is not arbitrary: it is induced by the structural coordinates defined in Theorem~\ref{thm:coarsest}. How small can $\|D^{(z)}\|_1$ actually be under this construction? Since $B_0(z,x)=\delta_{zx}$, the reference configuration $z$ necessarily forms a singleton mixer class. Together with the vanishing row and column sums of $D^{(z)}$, this property yields the following lower bound.

\begin{proposition}[Minimal dependence scale]
For any reference configuration $z$,
\begin{equation}
    \|D^{(z)}\|_1
\geq \frac{4[1-P_J^{(z)}(c(z))]}{N}.
    \label{eq:minimal_dependence}
\end{equation}
In particular, if $P_J^{(z)}(c(z))\leq 1-\varepsilon$ for some constant $\varepsilon>0$ independent of $N$, then $\|D^{(z)}\|_1=\Omega(N^{-1})$.
\end{proposition}
The condition $P_J^{(z)}(c(z))\leq 1-\varepsilon$ is mild, requiring only that the probability mass of the cost class containing $z$ remain bounded away from one. Combining the preceding two propositions, we can establish a bound on the total QAOA amplitude and derive necessary conditions for exponential amplification.

\begin{theorem}[Necessary dependence for exponential amplification]
Consider a family of single-layer QAOA instances initialized in the uniform superposition $|s\rangle$, with an unbiased mixer (Definition~\ref{def:unbiased}) and reference configurations $z$, such that $P_J^{(z)}(c(z))\leq 1-\varepsilon$ for some constant $\varepsilon>0$ independent of $N$. Then,
\begin{equation}
    |\psi_z|=O\left(\sqrt{N}\,\|D^{(z)}\|_1\right).
    \label{eq:total_dependence_bound}
\end{equation}
Consequently, achieving $|\psi_z|=\Omega(N^{-1/2+\delta})$ for some fixed $\delta>0$ necessarily requires
\begin{equation}
    \|D^{(z)}\|_1=\Omega(N^{-1+\delta}).
    \label{eq:necessary_dependence}
\end{equation}
\end{theorem}

The question of amplification beyond the Grover scale is naturally formulated in terms of QAOA with arbitrary $p$. At $p=1$, however, the Grover benchmark corresponds to a constant-factor improvement over the uniform amplitude $N^{-1/2}$. More generally, we are interested in whether single-layer QAOA can achieve an exponential amplification relative to this scale. According to the preceding theorem, achieving an amplitude of order $N^{-1/2+\varepsilon}$, for any fixed $\varepsilon>0$, necessarily requires $\|D^{(z)}\|_1=\Omega(N^{-1+\varepsilon})$. Thus, exponential amplification requires statistical dependence that is polynomially larger in $N$ than the minimal $N^{-1}$ scale. Although it remains unclear how these results extend to arbitrary depth $p$, they suggest that statistical dependence between the problem and mixer coordinates may play a fundamental role in enabling amplification beyond the Grover scale.

We now return to the Grover mixer discussion and examine how its limited structural resolution constrains statistical dependence. By comparing unstructured search with arbitrary mixers and the Grover mixer with arbitrary cost functions, we show that either a structurally simple problem or a mixer with limited structural resolution can restrict the dependence to the minimal scale $\|D^{(z)}\|_1=\Theta(N^{-1})$.

\paragraph{Grover mixer with unstructured search.} Consider a unique marked configuration $z$, with $c(z)=0$ and $c(x)=1$ for $x\neq z$. The Grover mixer distinguishes only $z$ from all other configurations, so $J$ and $K$ are both binary, with joint distribution
\begin{equation}
D^{(z)}=\left(\frac{1}{N}-\frac{1}{N^2}\right)
\begin{pmatrix}
1 & -1\\
-1 & 1
\end{pmatrix}.
\end{equation}
Consequently,
\begin{equation}
\|D^{(z)}\|_1=4\left(\frac{1}{N}-\frac{1}{N^2}\right)=\Theta(N^{-1}).
\end{equation}

\paragraph{Unstructured search with an arbitrary mixer.} Consider the same cost function, but allow an arbitrary mixer geometry satisfying the structural construction of Theorem~\ref{thm:coarsest}. Since $z$ is the unique configuration with $J=0$, and its joint class is a singleton, the dependence function satisfies
\begin{equation}
    D^{(z)}(0,K_z)=\frac{1}{N}-\frac{1}{N^2},
    \qquad
    D^{(z)}(0,k)=-\frac{P_K^{(z)}(k)}{N},\quad k\neq K_z,
\end{equation}
with $D^{(z)}(1,k)=-D^{(z)}(0,k)$. Summing the absolute values gives
\begin{equation}
    \|D^{(z)}\|_1=4\left(\frac{1}{N}-\frac{1}{N^2}\right)=\Theta(N^{-1}).
\end{equation}
Thus, the minimal dependence scale is attained independently of the mixer geometry.

\paragraph{Grover mixer with an arbitrary cost function.} Conversely, consider an arbitrary cost function with the Grover mixer. Since the mixer distinguishes only the reference configuration $z$ from its complement, we have
\begin{equation}
    D^{(z)}(j,K_z)=\frac{\delta_{j,c(z)}-P_J^{(z)}(j)}{N},
    \qquad
    D^{(z)}(j,k)=-D^{(z)}(j,K_z),\quad k\neq K_z.
\end{equation}
Therefore,
\begin{equation}
    \|D^{(z)}\|_1=\frac{4[1-P_J^{(z)}(c(z))]}{N}.
\end{equation}
In particular, if $P_J^{(z)}(c(z))\leq 1-\varepsilon$ for some constant $\varepsilon>0$, then $\|D^{(z)}\|_1=\Theta(N^{-1})$.

These examples show that minimal dependence can arise either from the limited structure of the cost function or from the limited resolution of the mixer coordinate. In both cases, the dependence remains at the $N^{-1}$ scale, preventing exponential amplitude amplification for unbiased mixers.

These results establish statistical dependence between the problem and mixer coordinates as a necessary resource for amplitude amplification. However, the magnitude of this dependence alone does not determine how effectively QAOA can exploit it, since substantial dependence need not translate into constructive interference. The achievable amplification also depends on the variational parameters, which control interference between problem and mixer contributions. Thus, the structural framework identifies a necessary resource for amplification, but does not yet determine how much of it can be exploited.

\section{How much can optimization improve QAOA performance?}
\label{sec:optimization}

A second fundamental aspect of QAOA performance concerns the role of variational optimization. Understanding the limits of parameter optimization provides an essential piece for addressing the limitations of the structural characterization developed in the preceding section, a connection we will revisit in the next section. Beyond this connection, determining the extent to which optimization can improve measurement probabilities is a fundamental question in its own right, with important algorithmic consequences that we explore in this section.

\subsection{Polynomial mean--max proximity}
\label{subsec:mean-max}

One way to assess the potential improvement achievable through variational optimization is to compare the optimal measurement probability with its average over the parameter space. This average provides a natural baseline for typical performance, allowing us to investigate the extent to which optimization can improve upon it. Unlike the structural analyses in the preceding and following sections, our approach here allows for a more general setting, encompassing arbitrary initial states, subsets of measurement outcomes, and arbitrary circuit depth $p$, rather than being restricted to a uniform initial superposition, a single reference configuration, and a single QAOA layer. In the following theorem, we establish a general relationship between the optimal and average measurement probabilities, revealing a remarkable constraint on the improvement achievable through variational optimization.

\begin{theorem}[Polynomial Mean--max proximity]
\label{thm:mean-max}
Consider depth-$p$ QAOA initialized in an arbitrary normalized state $|\psi_0\rangle$, with a real-valued cost function $c$ taking $m$ distinct values and an arbitrary Hermitian mixer Hamiltonian $H_M$ with $d$ distinct eigenvalues. For any subset $A\subseteq\mathcal{S}$,
\begin{equation}
\label{eq:mean-max}
\overline{q}_A^{(p)}
\leq
\sup_{\boldsymbol{\gamma},\boldsymbol{\beta}\in\mathbb{R}^p}
q_A^{(p)}(\boldsymbol{\gamma},\boldsymbol{\beta})
\leq
(md)^p\,\overline{q}_A^{(p)}.
\end{equation}
In particular, for a family of problem instances associated with an NPO-PB problem of size $n$, using a mixer Hamiltonian with a polynomially bounded number of distinct eigenvalues, $d=\operatorname{poly}(n)$, and further assuming that $md=n^{\Theta(1)}$, the factor $(md)^p$ is polynomially bounded if and only if $p=O(1)$. Consequently, at any constant depth,
\begin{equation}
\label{eq:mean-max-poly}
\sup_{\boldsymbol{\gamma},\boldsymbol{\beta}\in\mathbb{R}^p}q_A^{(p)}(\boldsymbol{\gamma},\boldsymbol{\beta})\overset{\mathrm{poly}}{\asymp}\overline{q}_A^{(p)}.
\footnote{Here, $a_n\overset{\mathrm{poly}}{\asymp}b_n$ means that there exists a polynomial $p(n)>0$ such that $p(n)^{-1}b_n\leq a_n\leq p(n)b_n$ for all sufficiently large $n$.}
\end{equation}
\end{theorem}

\begin{proof}
Let $\mathcal J$ and $\Lambda$ denote the sets of distinct eigenvalues of $H_C$ and $H_M$, respectively, with $|\mathcal J|=m$ and $|\Lambda|=d$. Their spectral decompositions are
\begin{equation}
H_C=\sum_{j\in\mathcal J}jQ_j,
\qquad
H_M=\sum_{\lambda\in\Lambda}\lambda\Pi_\lambda,
\end{equation}
where $Q_j$ and $\Pi_\lambda$ are the corresponding spectral projectors. Consequently, the depth-$p$ amplitude admits the expansion
\begin{equation}
\psi_z^{(p)}(\boldsymbol{\gamma},\boldsymbol{\beta})
=
\sum_{\mathbf j\in\mathcal J^p}
\sum_{\boldsymbol{\lambda}\in\Lambda^p}
a_{\mathbf j,\boldsymbol{\lambda}}^{(z)}
e^{-i(\boldsymbol{\gamma}\cdot\mathbf j+
\boldsymbol{\beta}\cdot\boldsymbol{\lambda})}, \qquad
a_{\mathbf j,\boldsymbol{\lambda}}^{(z)}
=
\langle z|
\Pi_{\lambda_p}Q_{j_p}
\cdots
\Pi_{\lambda_1}Q_{j_1}
|\psi_0\rangle.
\end{equation}
Applying a Parseval-type identity for finite exponential sums to all $2p$ parameters yields
\begin{equation}
\overline{q}_z^{(p)}
=
\lim_{T_\gamma,T_\beta\to\infty}
\frac{1}{T_\gamma^p T_\beta^p}
\int_{[0,T_\gamma]^p}
\int_{[0,T_\beta]^p}
q_z^{(p)}(\boldsymbol{\gamma},\boldsymbol{\beta})
\,d^p\boldsymbol{\beta}\,d^p\boldsymbol{\gamma}
=
\sum_{\mathbf j\in\mathcal J^p}
\sum_{\boldsymbol{\lambda}\in\Lambda^p}
\left|a_{\mathbf j,\boldsymbol{\lambda}}^{(z)}\right|^2.
\end{equation}
Here, the equality follows from the asymptotic orthogonality of exponentials associated with distinct eigenvalues of each Hamiltonian. By the Cauchy--Schwarz inequality,
\begin{equation}
\begin{aligned}
q_z^{(p)}(\boldsymbol{\gamma},\boldsymbol{\beta})
&\leq
\left(
\sum_{\mathbf j,\boldsymbol{\lambda}}
\left|a_{\mathbf j,\boldsymbol{\lambda}}^{(z)}\right|^2
\right)
\left(
\sum_{\mathbf j,\boldsymbol{\lambda}}
\left|e^{-i(\boldsymbol{\gamma}\cdot\mathbf j+
\boldsymbol{\beta}\cdot\boldsymbol{\lambda})}\right|^2
\right)
\\
&=
(md)^p\,\overline{q}_z^{(p)}.
\end{aligned}
\end{equation}
Now, for any subset $A\subseteq\mathcal S$, summing the pointwise bound over $z\in A$, we obtain
\begin{equation}
q_A^{(p)}(\boldsymbol{\gamma},\boldsymbol{\beta})
\leq
(md)^p\sum_{z\in A}\overline{q}_z^{(p)}
=
(md)^p\,\overline{q}_A^{(p)}.
\end{equation}
Since the supremum is at least as large as the average, this establishes Eq.~\eqref{eq:mean-max}.

For Eq.~\eqref{eq:mean-max-poly}, observe that the NPO-PB condition implies that the number of distinct cost values is polynomially bounded, $m=\operatorname{poly}(n)$. Together with $d=\operatorname{poly}(n)$ and the additional assumption $md=n^{\Theta(1)}$, we have $(md)^p=n^{\Theta(p)}$, which is polynomially bounded if and only if $p=O(1)$. At constant depth, the claimed polynomial equivalence follows immediately.
\end{proof}

Under the polynomial boundedness assumptions, Theorem~\ref{thm:mean-max} establishes that the average and optimal measurement probabilities have the same superpolynomial scaling classification, since they differ by at most polynomial factors. Exponential scaling, which is often of particular interest, is included as a special case. These conditions are satisfied by several canonical QAOA settings. On the mixer side, both the transverse field and Grover mixers have a polynomially bounded number of distinct eigenvalues. On the problem side, the NPO-PB class includes canonical problems such as Max-Cut, Max-SAT, and Maximum Independent Set in their unweighted formulations. This class is particularly relevant because the polynomial boundedness of the objective function allows expectation values to be efficiently estimated through quantum sampling, as shown by Marsh and Wang~\cite{marsh2019quantum}.

Importantly, the mean--max relation constrains the maximum achievable probability, rather than the probability obtained for every choice of variational parameters. Particular choices may still lead to constructive or destructive interference, potentially even complete cancellation. Furthermore, this relation does not imply that the average or optimal probabilities can be computed efficiently in general.

The generalized mean--max bound also extends to arbitrary positive semidefinite observables. Indeed, for any $O\succeq 0$, the expectation value $\langle O\rangle^{(p)}=\langle\psi^{(p)}|O|\psi^{(p)}\rangle$ can be written as $\|O^{1/2}|\psi^{(p)}\rangle\|^2$. The same spectral decomposition and Cauchy--Schwarz argument therefore yield
\begin{equation}
\overline{\langle O\rangle}^{(p)}
\leq
\sup_{\boldsymbol{\gamma},\boldsymbol{\beta}}
\langle O\rangle^{(p)}
\leq
(md)^p\,\overline{\langle O\rangle}^{(p)}.
\end{equation}
This bound applies directly to the expected cost $\langle H_C\rangle^{(p)}=\sum_{z\in\mathcal{S}}c(z)q_z^{(p)}$ whenever $c(z)\geq 0$ for all $z\in\mathcal{S}$. This nonnegativity condition is not restrictive, since any cost function can be shifted by a constant to make all its values nonnegative without changing the optimal solutions. For NPO-PB problems, however, this extension is generally of limited interest, since the expectation value is already bounded by the polynomial range of the objective function, unless its parameter average is superpolynomially small, which would be highly atypical in standard optimization settings under a natural normalization of the objective function. In contrast, measurement probabilities may be superpolynomially small, so the corresponding mean--max bound is not rendered trivial by the range of the quantity being considered. Nevertheless, as discussed below, the mean--max bound still has important implications for the optimization of observables such as the expectation value.

\subsection{Algorithmic implications of mean--max proximity}
\label{subsec:alg_imp}

The polynomial proximity between average and optimal success probabilities suggests potentially far-reaching algorithmic implications, particularly regarding the extent to which variational optimization is necessary for achieving high success probabilities in QAOA. In this subsection, we explore and formalize these implications by investigating the consequences of polynomial mean--max proximity for random-parameter sampling and characterizing limitations of variational optimization in QAOA.

\subsubsection{Anticoncentration}
\label{subsec:anticoncentration}

The mean--max result not only establishes polynomial proximity between average and optimal probabilities but also implies that parameter choices achieving a constant fraction of the average probability cannot be superpolynomially rare. The following corollary quantifies this anticoncentration property. Throughout the remainder of this section, we assume that QAOA probabilities are periodic in each variational parameter.

\begin{corollary}[Anticoncentration of QAOA measurement probabilities]
\label{cor:anticoncentration}
Consider depth-$p$ QAOA initialized in an arbitrary normalized state $|\psi_0\rangle$, with a real-valued cost function $c$ taking $m$ distinct values and an arbitrary Hermitian mixer Hamiltonian $H_M$ with $d$ distinct eigenvalues. Assume that the QAOA probabilities are periodic in each variational parameter and that the parameters are independently and uniformly distributed over one full period of each parameter. For any subset $A\subseteq\mathcal{S}$ with $\overline{q}_A^{(p)}>0$ and any fixed constant $\alpha\in(0,1)$,
\begin{equation}
\label{eq:anticoncentration}
\Pr_{\boldsymbol{\gamma},\boldsymbol{\beta}}
\left(
q_A^{(p)}(\boldsymbol{\gamma},\boldsymbol{\beta})
\geq
\alpha\overline{q}_A^{(p)}
\right)
\geq
\frac{1-\alpha}{(md)^p-\alpha}.
\end{equation}
In particular, for a family of problem instances associated with an NPO-PB problem of size $n$, using a mixer Hamiltonian with a polynomially bounded number of distinct eigenvalues, $d=\operatorname{poly}(n)$, at any constant depth $p=O(1)$, we have
\begin{equation}
\label{eq:anticoncentration-poly}
\Pr_{\boldsymbol{\gamma},\boldsymbol{\beta}}
\left(
q_A^{(p)}(\boldsymbol{\gamma},\boldsymbol{\beta})
\geq
\alpha\overline{q}_A^{(p)}
\right)
\geq
\frac{1}{\operatorname{poly}(n)}.
\end{equation}
\end{corollary}

\begin{proof}
Let $X=q_A^{(p)}(\boldsymbol{\gamma},\boldsymbol{\beta})$, $\mu=\overline{q}_A^{(p)}>0$, and $P=\Pr(X\geq\alpha\mu)$. By Theorem~\ref{thm:mean-max}, $0\leq X\leq(md)^p\mu$. Uniform averaging over one full period of each parameter gives $\mathbb{E}[X]=\mu$. Thus, $\mu=\mathbb{E}[X]\leq\alpha\mu(1-P)+(md)^p\mu P$, which, upon rearranging, gives Eq.~\eqref{eq:anticoncentration}. The inverse-polynomial bound in Eq.~\eqref{eq:anticoncentration-poly} follows directly from $m=\operatorname{poly}(n)$, $d=\operatorname{poly}(n)$, and $p=O(1)$.
\end{proof}

\subsubsection{Random-parameter QAOA}

The anticoncentration result established above shows that, under polynomial mean--max proximity, parameter choices yielding measurement probabilities comparable to the average are not exceedingly rare. This raises a natural question: can random parameter sampling be used directly to obtain good solutions, without identifying or optimizing individual parameter choices? To investigate this, we consider a simple QAOA algorithm that independently samples parameters at each execution and retains the measured configuration with the best objective value. The procedure is formalized in Algorithm~\ref{alg:random-parameter}.

\begin{algorithm}[H]
\caption{Random-parameter QAOA}
\label{alg:random-parameter}
\begin{algorithmic}[1]
\Require Problem Hamiltonian $H_C$, mixer Hamiltonian $H_M$, depth $p$, number of executions $R$
\State Let $\boldsymbol{\theta}=(\boldsymbol{\gamma},\boldsymbol{\beta})$ denote the complete QAOA parameter vector
\For{$j=1,\ldots,R$}
    \State Sample $\boldsymbol{\theta}_j$ independently from the parameter distribution
    \State Prepare the QAOA state $|\psi_p(\boldsymbol{\theta}_j)\rangle$
    \State Measure in the computational basis, obtaining $z_j$
\EndFor
\State \Return $z^*=\arg\max_{j=1,\ldots,R} c(z_j)$
\end{algorithmic}
\end{algorithm}

Before proceeding, we clarify that our analysis will not necessarily focus on the final configuration returned by the algorithm, but may also consider the configurations sampled during its execution. In this sense, for any subset $A\subseteq\mathcal{S}$, let $P_A^{\mathrm{rand}}(R)$ denote the probability that at least one of the $R$ independent random-parameter executions yields a configuration in $A$. In particular, observe that when $A$ is defined by an objective-value threshold, i.e., $A=\{z\in\mathcal{S}:c(z)\geq c_0\}$, this probability coincides with the probability that the algorithm returns a configuration in $A$. The following theorem provides a lower bound on $P_A^{\mathrm{rand}}(R)$ in terms of the maximum success probability achievable by QAOA.

\begin{theorem}[Random-parameter QAOA guarantee]
\label{thm:random-qaoa}
Consider depth-$p$ QAOA initialized in an arbitrary normalized state $|\psi_0\rangle$, with a real-valued cost function $c$ taking $m$ distinct values and an arbitrary Hermitian mixer Hamiltonian $H_M$ with $d$ distinct eigenvalues. Assume that the QAOA probabilities are periodic in each variational parameter and that the parameters are independently and uniformly distributed over one full period of each parameter. For any subset $A\subseteq\mathcal{S}$, the probability that at least one of the $R=(md)^p$ independent samples generated by Algorithm~\ref{alg:random-parameter} belongs to $A$ satisfies
\begin{equation}
\label{eq:random-qaoa-guarantee}
P_A^{\mathrm{rand}}(R)\geq (1-e^{-1})\sup_{\boldsymbol{\gamma},\boldsymbol{\beta}\in\mathbb{R}^p}q_A^{(p)}(\boldsymbol{\gamma},\boldsymbol{\beta}).
\end{equation}
In particular, for a family of problem instances associated with an NPO-PB problem of size $n$, using a mixer Hamiltonian with a polynomially bounded number of distinct eigenvalues, $d=\operatorname{poly}(n)$, at any constant depth $p=O(1)$, the number of executions $R=(md)^p$ is polynomially bounded. Assuming that the initial state can be efficiently prepared, the mixer Hamiltonian $H_M$ can be efficiently implemented, and the QAOA parameters can be efficiently sampled, Algorithm~\ref{alg:random-parameter} can be executed in polynomial time, with the probability that at least one of its measured configurations belongs to $A$ satisfying the guarantee in Eq.~\eqref{eq:random-qaoa-guarantee}.
\end{theorem}

\begin{proof}
Since each execution independently samples QAOA parameters and performs one measurement, the probability of obtaining a configuration in $A$ is $\overline q_A^{(p)}$. The probability that at least one of the $R$ executions produces a configuration in $A$ is consequently
\begin{equation}
P_A^{\mathrm{rand}}(R)=1-\left(1-\overline q_A^{(p)}\right)^R.
\end{equation}
Using $1-x\leq e^{-x}$, setting $R=(md)^p$, and applying Theorem~\ref{thm:mean-max}, we obtain
\begin{equation}
P_A^{\mathrm{rand}}(R)\geq 1-e^{-(md)^p\overline q_A^{(p)}}\geq 1-e^{-\sup_{\boldsymbol{\gamma},\boldsymbol{\beta}\in\mathbb{R}^p}q_A^{(p)}(\boldsymbol{\gamma},\boldsymbol{\beta})}.
\end{equation}
Thus, since $0\leq\sup_{\boldsymbol{\gamma},\boldsymbol{\beta}\in\mathbb{R}^p}q_A^{(p)}(\boldsymbol{\gamma},\boldsymbol{\beta})\leq 1$ and $1-e^{-x}\geq(1-e^{-1})x$ for $x\in[0,1]$, Eq.~\eqref{eq:random-qaoa-guarantee} follows.

To finish, under the stated efficiency assumptions, each execution takes polynomial time. Note that NPO requires the objective function to be computable in polynomial time, which allows the diagonal Hamiltonian $H_C$ to be efficiently implemented by Rule 1.6 of Ref.~\cite{childs}. Furthermore, since $m,d=\operatorname{poly}(n)$ and $p=O(1)$, we have $R=(md)^p=\operatorname{poly}(n)$. Therefore, Algorithm~\ref{alg:random-parameter} can be executed in polynomial time while satisfying the guarantee in Eq.~\eqref{eq:random-qaoa-guarantee}.
\end{proof}

Under the conditions of the theorem, random-parameter QAOA achieves at least a fraction $1-e^{-1}\approx 0.632$ of the optimal single-execution success probability using $(md)^p$ executions, without variational optimization.

\subsubsection{Implications for variational optimization}
\label{subsec:variational-optimization}

The preceding result naturally raises the question of how random-parameter QAOA compares with the standard variational procedure, in which parameters are optimized with respect to an observable before sampling candidate solutions. The latter is formalized in Algorithm~\ref{alg:variational-qaoa}, with the optimization stage treated independently of the particular classical method employed.

\begin{algorithm}[H]
\caption{Variational QAOA}
\label{alg:variational-qaoa}
\begin{algorithmic}[1]
\Require Cost function $c$, QAOA circuit of depth $p$, observable $O$, number of executions $V$
\State Obtain $(\boldsymbol{\gamma}_O^*,\boldsymbol{\beta}_O^*)$, where the subscript $O$ denotes the observable whose expectation value $\langle O\rangle$ is variationally optimized
\For{$j=1,\ldots,V$}
    \State Prepare $|\psi_p(\boldsymbol{\gamma}_O^*,\boldsymbol{\beta}_O^*)\rangle$
    \State Measure in the computational basis, obtaining $z_j$
\EndFor
\State \Return $z^*=\arg\max_{j=1,\ldots,V} c(z_j)$
\end{algorithmic}
\end{algorithm}

Let $P_A^{\mathrm{var}}(V)$ denote the probability that at least one of $V$ independent executions of the variationally optimized QAOA circuit produces a configuration in $A$. We now compare the cumulative success probabilities of Algorithms~\ref{alg:random-parameter} and~\ref{alg:variational-qaoa} for any subset $A\subseteq\mathcal{S}$. The following corollary of Theorem~\ref{thm:random-qaoa} establishes a quantitative guarantee relating their success probabilities.

\begin{corollary}[Random-parameter versus variational QAOA]
\label{cor:random-vs-variational}
Consider depth-$p$ QAOA initialized in an arbitrary normalized state $|\psi_0\rangle$, with a real-valued cost function $c$ taking $m$ distinct values and an arbitrary Hermitian mixer Hamiltonian $H_M$ with $d$ distinct eigenvalues. Assume that the QAOA probabilities are periodic in each variational parameter and that the parameters are independently and uniformly distributed over one full period of each parameter. For any subset $A\subseteq\mathcal{S}$, let $(\boldsymbol{\gamma}_O^*,\boldsymbol{\beta}_O^*)$ be the parameters obtained from Variational QAOA (Algorithm~\ref{alg:variational-qaoa}). For any positive integer $V$, Random-parameter QAOA (Algorithm~\ref{alg:random-parameter}), with $R=(md)^pV$ independent executions, satisfies
\begin{equation}
\label{eq:random-vs-variational}
P_A^{\mathrm{rand}}(R)\geq (1-e^{-1})P_A^{\mathrm{var}}(V).
\end{equation}
\end{corollary}

\begin{proof}
We divide the $R=(md)^pV$ independent random-parameter executions into $V$ independent blocks, each containing $(md)^p$ executions. Let $q_{\max}=\sup_{\boldsymbol{\gamma},\boldsymbol{\beta}\in\mathbb{R}^p}q_A^{(p)}(\boldsymbol{\gamma},\boldsymbol{\beta})$ and $q_*=q_A^{(p)}(\boldsymbol{\gamma}_O^*,\boldsymbol{\beta}_O^*)$. By Theorem~\ref{thm:random-qaoa}, the probability that at least one execution in a given block produces a configuration in $A$ is at least $(1-e^{-1})q_{\max}\geq(1-e^{-1})q_*$. Since the blocks are independent, we obtain
\begin{equation}
P_A^{\mathrm{rand}}(R)\geq 1-\left[1-(1-e^{-1})q_*\right]^V\geq(1-e^{-1})\left[1-(1-q_*)^V\right]=(1-e^{-1})P_A^{\mathrm{var}}(V),
\end{equation}
where the second inequality follows from the concavity of $f(x)=1-(1-x)^V$ on $[0,1]$ and $f(0)=0$.
\end{proof}

Thus, with a multiplicative overhead of $(md)^p$ in circuit executions, random-parameter QAOA achieves at least a constant fraction $(1-e^{-1})$ of the cumulative success probability of $V$ executions of the trained variational circuit, without requiring parameter optimization. When this overhead is polynomially bounded, the comparison yields a further consequence for the computational efficiency of random-parameter sampling.

\begin{corollary}[Efficient sampling without variational training]
\label{cor:training-free-sampling}
Consider depth-$p$ QAOA initialized in an arbitrary normalized state $|\psi_0\rangle$, with a real-valued cost function $c$ taking $m$ distinct values and an arbitrary Hermitian mixer Hamiltonian $H_M$ with $d$ distinct eigenvalues. Assume that the QAOA probabilities are periodic in each variational parameter and that the parameters are independently and uniformly distributed over one full period of each parameter. For a family of problem instances associated with an NPO-PB problem of size $n$, using a mixer Hamiltonian with a polynomially bounded number of distinct eigenvalues, $d=\operatorname{poly}(n)$, at any constant depth $p=O(1)$, consider any family of reference sets $A\subseteq\mathcal{S}$ and let $(\boldsymbol{\gamma}_O^*,\boldsymbol{\beta}_O^*)$ be the parameters obtained from Variational QAOA (Algorithm~\ref{alg:variational-qaoa}), satisfying
\begin{equation}
\label{eq:variational-efficient-success}
q_A^{(p)}(\boldsymbol{\gamma}_O^*,\boldsymbol{\beta}_O^*)\geq\frac{1}{\operatorname{poly}(n)}.
\end{equation}
Then, assuming that the initial state can be efficiently prepared, the mixer Hamiltonian $H_M$ can be efficiently implemented, and the QAOA parameters can be efficiently sampled, Random-parameter QAOA (Algorithm~\ref{alg:random-parameter}) samples at least one configuration in $A$ with constant probability using polynomially many independent executions.
\end{corollary}

\begin{proof}
By assumption, $q_*=q_A^{(p)}(\boldsymbol{\gamma}_O^*,\boldsymbol{\beta}_O^*)\geq 1/h(n)$ for some positive polynomial $h(n)$. Since
\begin{equation}
P_A^{\mathrm{var}}(V)=1-(1-q_*)^V\geq 1-e^{-V/h(n)},
\end{equation}
choosing $V=\lceil h(n)\ln n\rceil$ yields $P_A^{\mathrm{var}}(V)\geq 1-1/n$.

By Corollary~\ref{cor:random-vs-variational}, for $R=(md)^pV$ independent random-parameter executions,
\begin{equation}
P_A^{\mathrm{rand}}(R)\geq(1-e^{-1})P_A^{\mathrm{var}}(V)\geq(1-e^{-1})(1-1/n).
\end{equation}
Thus, for $n\geq 2$, we have $P_A^{\mathrm{rand}}(R)\geq (1-e^{-1})/2>0$. Since $(md)^p=\operatorname{poly}(n)$ and $V=O(h(n)\log n)$, the total number of executions $R=(md)^pV$ is polynomial in $n$. Under the stated efficiency assumptions, each circuit execution takes polynomial time. Therefore, Random-parameter QAOA achieves the claimed constant success probability in polynomial time.
\end{proof}

The corollary shows that whenever a variationally optimized QAOA circuit achieves an inverse-polynomial success probability, random-parameter QAOA can also sample configurations in the reference set with constant probability using polynomially many circuit executions, without variational training and regardless of the observable used during optimization. Importantly, the result does not require the variational parameters to maximize the success probability, but only to achieve an inverse-polynomial value, and therefore also applies to suboptimal parameters. This establishes that efficient sampling does not necessarily require finding suitable variational parameters, provided that the overhead is polynomially bounded.

It is worth noting that previous results, such as those of Ref.~\cite{bittel2021training}, have established that training variational quantum algorithms, including QAOA, is NP-hard in the worst case. This does not contradict our findings, since the computational difficulty of variational optimization does not necessarily imply the difficulty of sampling desirable configurations. Our result does not establish that optimal variational parameters can be found efficiently; rather, it shows that whenever parameters achieving inverse-polynomial success probability exist, random-parameter QAOA can sample desirable configurations with constant probability using polynomially many circuit executions under polynomial mean--max proximity, without explicitly finding those parameters.

\section{The pre-optimization structure of QAOA performance}
\label{sec:pre-optimization}

We now use the mean--max relation established in Section~\ref{sec:optimization} to revisit the structural characterization developed in Section~\ref{sec:structure}, investigating how problem structure contributes to amplitude amplification in the setting where polynomial mean--max proximity holds. Specifically, we return to single-layer QAOA, with a uniform superposition initial state and a single reference state amplitude, for which the structural approach is particularly convenient. In this setting, the general mean--max relation established in Theorem~\ref{thm:mean-max} then yields the following corollary.

\begin{corollary}[Single-layer mean--max proximity]
\label{cor:p-1-mean-max}
Consider single-layer QAOA initialized in the uniform superposition $|s\rangle$, with a real-valued cost function $c$ taking $m$ distinct values and an arbitrary Hermitian mixer Hamiltonian $H_M$ with $d$ distinct eigenvalues. For any reference configuration $z$,
\begin{equation}
\label{eq:p-1-mean-max}
\overline{q}_z
\leq
\sup_{\gamma,\beta\in\mathbb R}
q_z(\gamma,\beta)
\leq
md\,\overline{q}_z.
\end{equation}
In particular, for a family of problem instances associated with an NPO-PB problem of size $n$, using a mixer Hamiltonian with a polynomially bounded number of distinct eigenvalues, $d=\operatorname{poly}(n)$, we have
\begin{equation}
\label{eq:p-1-mean-max-poly}
\sup_{\gamma,\beta\in\mathbb R}q_z(\gamma,\beta)
\overset{\mathrm{poly}}{\asymp}
\overline{q}_z.
\end{equation}
\end{corollary}

Throughout this section, we work in the regime of polynomial mean--max proximity established in Corollary~\ref{cor:p-1-mean-max}. In particular, we adopt the following assumptions to guide our asymptotic analysis, while stating each formal result under its own minimal hypotheses. We consider families of NPO-PB problem instances of size $n$ (implying $m=\operatorname{poly}(n)$), with mixer Hamiltonians satisfying $d=\operatorname{poly}(n)$ and $\ell=\operatorname{poly}(n)$. The latter condition is an additional assumption for the structural analysis and is not required for polynomial mean--max proximity.

\subsection{QAOA performance as a structural overlap}

As discussed in Section~\ref{sec:structure}, a central difficulty in characterizing amplitude amplification in QAOA is its dependence on the variational parameters, which control the interference between problem and mixer contributions. The dependence-based framework developed there identifies a necessary structural resource for amplification, but does not determine how effectively this resource can be exploited by the algorithm. Given the polynomial mean--max proximity established above, we can instead study the asymptotic scaling of the optimal measurement probability, up to polynomial factors, through its parameter average. Averaging removes the explicit parameter dependence, potentially making the problem more analytically tractable and allowing us to seek a structural characterization of the resulting probability.

This connection reveals that the asymptotic scaling of the optimal reference state measurement probability is already determined, up to polynomial factors, by a pre-optimization structural layer. This suggests two complementary interpretations of the role of variational optimization. On the one hand, there is always a choice of variational parameters that achieves at least the average measurement probability. On the other hand, optimization cannot improve upon this average by more than a polynomial factor. It is worth emphasizing that this does not diminish the role of quantum interference, which remains the physical mechanism underlying the actual amplification of the reference state amplitude. What changes is that we can now characterize the asymptotic scaling of the optimal probability without explicitly identifying the parameters that produce favorable interference. It is worth emphasizing that the existence of a pre-optimization structural layer is a general consequence of polynomial mean--max proximity, rather than a property restricted to single-layer QAOA. The setting considered here provides a particular case in which this layer can be explicitly characterized.

Our next step is to characterize this pre-optimization structural layer explicitly. In particular, we seek to understand how the interplay between problem structure and QAOA dynamics determines the average measurement probability.

To establish this characterization, it is convenient to separate the distribution of problem costs within each mixer class from the relative weight of that class. Specifically, we introduce the conditional probability $P^{(z)}(J\mid k)$, which describes the distribution of problem classes within mixer class $k$, while $P_K^{(z)}(k)$ accounts for its relative weight. Using $P^{(z)}(j,k)=P^{(z)}(j\mid k)P_K^{(z)}(k)$, we can rewrite Eq.~\eqref{eq:amplitude} as
\begin{equation} \label{eq:conditional_amplitude}
\psi_z(\gamma,\beta)
=
\sqrt{N}\sum_{j,k}
P^{(z)}(j\mid k)P_K^{(z)}(k)B_\beta^{(z)}(k)e^{-i\gamma j}.
\end{equation}
The following theorem provides an exact structural characterization of the average measurement probability.

\begin{theorem}[Structural characterization of the average probability]
\label{thm:structural-average}
Consider single-layer QAOA initialized in the uniform superposition $|s\rangle$, with a real-valued cost function $c$ taking $m$ distinct values and an arbitrary Hermitian mixer Hamiltonian $H_M$ with $d$ distinct eigenvalues. For any reference configuration $z$,
\begin{equation} \label{eq:overlap}
\overline q_z
=
N\langle G_P^{(z)},M^{(z)}\rangle_{\mathrm{HS}} = N \sum_{a,b}(G_P^{(z)})_{ab}M_{ab}^{(z)},
\end{equation}
where $\langle A,B\rangle_{\mathrm{HS}}=\operatorname{Tr}(A^\dagger B)$ denotes the Hilbert--Schmidt inner product, and $G_P^{(z)},M^{(z)}\in\mathbb{R}^{\ell\times\ell}$ are defined by
\begin{equation}
G_P^{(z)}=(P^{(z)}_{J\mid K})^{\mathsf T}P^{(z)}_{J\mid K},
\qquad
M^{(z)}=P_K^{(z)} \operatorname{Re}\left\{
\,\mathbb E_\beta[B_\beta^{(z)} (B_\beta^{(z)})^\dagger]\,
\right\} P_K^{(z)}.
\end{equation}
Here, $[P^{(z)}_{J\mid K}]_{jk}=P^{(z)}(j\mid k)$, $P_K^{(z)}=\operatorname{diag}(P_K^{(z)}(k))$, and $[B_\beta^{(z)}]_k=B_\beta^{(z)}(k)$.

In particular, for a family of problem instances associated with an NPO-PB problem of size $n$, using a mixer Hamiltonian with a polynomially bounded number of distinct eigenvalues, $d=\operatorname{poly}(n)$, we have
\begin{equation} \label{eq:optimal_overlap}
\sup_{\gamma,\beta\in\mathbb R}q_z(\gamma,\beta)
\overset{\mathrm{poly}}{\asymp}
N\langle G_P^{(z)},M^{(z)}\rangle_{\mathrm{HS}}.
\end{equation}
\end{theorem}

\begin{proof}
Starting from Eq.~\eqref{eq:conditional_amplitude}, using the matrix notation introduced in the theorem, averaging over $\gamma$, and applying the definition of $G_P^{(z)}$, we obtain
\begin{equation}
\overline q_z(\beta)
=N(B_\beta^{(z)})^\dagger P_K^{(z)}G_P^{(z)}P_K^{(z)}B_\beta^{(z)}.
\end{equation}
Since the quadratic form is a scalar, we can express it as a trace. Averaging also over $\beta$ and using the cyclicity and linearity of the trace, we get
\begin{equation}
\overline q_z
=N\operatorname{Tr}\left(G_P^{(z)}P_K^{(z)}\mathbb E_\beta\left[B_\beta^{(z)}(B_\beta^{(z)})^\dagger\right]P_K^{(z)}\right).
\end{equation}
Since $G_P^{(z)}$ is real and symmetric and $P_K^{(z)}$ is real and diagonal, only the real part of $\mathbb E_\beta[B_\beta^{(z)}(B_\beta^{(z)})^\dagger]$ contributes to the trace. Therefore, $\overline q_z=N\operatorname{Tr}(G_P^{(z)}M^{(z)})$. 
The equivalences in Eq.~\eqref{eq:overlap} follow from the definition of the Hilbert--Schmidt inner product and the symmetry of $G_P^{(z)}$, while the summation form additionally uses the symmetry of $M^{(z)}$. The asymptotic equivalence in Eq.~\eqref{eq:optimal_overlap} then follows directly from Theorem~\ref{cor:p-1-mean-max}, completing the proof.
\end{proof}

\subsubsection{Properties of the Structural Overlap}

The matrices $G_P^{(z)}$ and $M^{(z)}$ admit complementary statistical interpretations and satisfy several general properties that will be useful in the subsequent analysis.

\begin{itemize}
    \item \textbf{Problem structure matrix ($G_P^{(z)}$).}

    By definition,
    \begin{equation}
    (G_P^{(z)})_{ab}
    =\sum_j P^{(z)}(j\mid a)P^{(z)}(j\mid b)
    =\langle P^{(z)}(J\mid a),P^{(z)}(J\mid b)\rangle.
    \end{equation}
    Thus, $G_P^{(z)}$ is the Gram matrix of the conditional cost distributions associated with the mixer classes.

    This matrix also admits a statistical interpretation. Let $J_a$ and $J_b$ be independent random variables distributed according to $P^{(z)}(J\mid K=a)$ and $P^{(z)}(J\mid K=b)$, respectively. Then
    \begin{equation}
    (G_P^{(z)})_{ab}=\Pr(J_a=J_b).
    \end{equation}
    Hence, the diagonal entries measure the concentration of costs within individual mixer classes, while the off-diagonal entries quantify the probability of cost coincidence between different classes.

    \item \textbf{Mixer accessibility matrix ($M^{(z)}$).}

    The matrix $M^{(z)}$ captures the dynamical correlations between mixer classes, weighted by their marginal probabilities. Defining $X_\beta^{(z)}=P_K^{(z)}B_\beta^{(z)}$, we can write
    \begin{equation}
    M^{(z)}=\operatorname{Re}\mathbb E_\beta[X_\beta^{(z)}(X_\beta^{(z)})^\dagger].
    \end{equation}
    Thus, $M^{(z)}$ is the real part of the uncentered second-moment matrix of $X_\beta^{(z)}$. Its diagonal entries give the average squared magnitudes of the weighted mixer amplitudes, while its off-diagonal entries capture their real-valued cross-correlations.
\end{itemize}

We can now summarize some general properties of $G_P^{(z)}$, $M^{(z)}$, and the structural overlap that will be useful in the subsequent analysis. Both $G_P^{(z)}$ and $M^{(z)}$ are real, symmetric, and positive semidefinite by construction. The diagonal elements $M_{aa}^{(z)}=P_K^{(z)}(a)^2\mathbb E_\beta[|B_\beta^{(z)}(a)|^2]$ are non-negative. Moreover, since $P^{(z)}(J\mid a)$ and $P^{(z)}(J\mid b)$ are probability distributions, the entries of $G_P^{(z)}$ satisfy $0\leq (G_P^{(z)})_{ab}\leq 1$.

The structural overlap also satisfies universal bounds. Since $\psi_z(0,0)=1/\sqrt{N}$, the mean--max relation in Corollary~\ref{cor:p-1-mean-max} implies $\overline q_z\geq 1/(mdN)$ for any reference configuration $z$. Combining this lower bound with the normalization $\overline q_z\leq 1$ and the structural representation $\overline q_z=N\langle G_P^{(z)},M^{(z)}\rangle_{\mathrm{HS}}$, we obtain
\begin{equation}
\label{eq:universal-overlap-bounds}
\frac{1}{mdN^2}
\leq
\langle G_P^{(z)},M^{(z)}\rangle_{\mathrm{HS}}
\leq
\frac{1}{N}.
\end{equation}
In particular, for NPO-PB problems with a polynomially bounded number of distinct mixer eigenvalues, $d=\operatorname{poly}(n)$, the structural overlap is bounded below by $1/[N^2\operatorname{poly}(n)]$. Thus, $N^{-2}$ is the smallest possible exponential scale of the structural overlap, up to polynomial factors.

The matrix $G_P^{(z)}$ is invariant under any injective transformation of the cost values. To see this, consider a transformed cost function $c'=f\circ c$, where $f$ is injective on the image of $c$. Since $c(x)=j$ if and only if $c'(x)=f(j)$, the conditional distributions satisfy $P_{c'}^{(z)}(f(j)\mid a)=P_c^{(z)}(j\mid a)$. Consequently,
\begin{equation}
\label{eq:GP-injective-invariance}
(G_{P,c'}^{(z)})_{ab}
=\sum_{j\in\operatorname{im}(c)}P_{c'}^{(z)}(f(j)\mid a)P_{c'}^{(z)}(f(j)\mid b)
=\sum_{j\in\operatorname{im}(c)}P_c^{(z)}(j\mid a)P_c^{(z)}(j\mid b)
=(G_{P,c}^{(z)})_{ab}.
\end{equation}
Thus, $G_P^{(z)}$ depends only on the distribution of equal-cost classes across mixer classes, not on the numerical values assigned to them. A notable special case is a permutation of the cost values. Since $M^{(z)}$ is also unaffected by such transformations, Eq.~\eqref{eq:overlap} implies that $\overline q_z$ remains invariant. In contrast, as can be seen from Eq.~\eqref{eq:amplitude}, injective transformations of the cost values generally change the QAOA probability for fixed choices of $\gamma$ and $\beta$, since they modify the phase factors $e^{-i\gamma j}$, even though the average probability remains unchanged.

Finally, the mapping from $P^{(z)}$ to $G_P^{(z)}$ is generally non-injective. For example, consider two distinct joint distributions with $N=8$:
\begin{equation}
P_1^{(z)}=\frac18
\begin{pmatrix}
1&1\\
0&0\\
0&3\\
0&3
\end{pmatrix},
\qquad
P_2^{(z)}=\frac18
\begin{pmatrix}
1&1\\
0&1\\
0&1\\
0&4
\end{pmatrix},
\qquad
G_{P,1}^{(z)}=G_{P,2}^{(z)}
=
\begin{pmatrix}
1&1/7\\
1/7&19/49
\end{pmatrix}.
\end{equation}
Thus, distinct joint distributions---not related by an injective transformation of the cost values---can yield the same problem structure matrix.

\subsubsection{Computing the mixer accessibility matrix}

Recall that $M^{(z)}$ is an intrinsic object associated with the mixer and the chosen reference vertex $z$, independent of the objective function. Observe that vertex transitivity provides a sufficient condition for independence from $z$, since graph automorphisms preserve walk counts and, consequently, the mixer classes and their transition amplitudes. This is the case for both the transverse field and Grover mixers.

It would therefore be useful to obtain a closed-form expression for $M^{(z)}$, allowing its contribution to the Structure--Accessibility Overlap to be evaluated without explicitly averaging over the variational parameter $\beta$. One possible general approach is to exploit the spectral decomposition of the mixer Hamiltonian. Writing the spectral decomposition as
\begin{equation}
H_M=\sum_\lambda\lambda\Pi_\lambda,
\qquad
U_M(\beta)=\sum_\lambda e^{-i\beta\lambda}\Pi_\lambda,
\end{equation}
where $\lambda$ runs over the distinct eigenvalues, it follows directly from the definition of $M^{(z)}$ that
\begin{equation}
M_{ab}^{(z)}
=
P_K^{(z)}(a)P_K^{(z)}(b)\operatorname{Re}\left\{
\sum_\lambda
\langle z|\Pi_\lambda|x_a\rangle
\langle x_b|\Pi_\lambda|z\rangle
\right\}.
\end{equation}
where $x_a$ and $x_b$ are representatives of mixer classes $a$ and $b$, respectively.

On the other hand, for mixers whose $B_\beta^{(z)}(k)$ admit closed-form expressions, it may be possible to compute $M^{(z)}$ analytically. This is the case for the transverse field and Grover mixers, as shown below.

\begin{itemize}
    \item \textbf{Transverse field mixer.}

    The mixer classes are characterized by the mixer factors in Eq.~\eqref{eq:TF_mixerfactors}, and their marginal probabilities are given by $P_K^{(z)}(a)=2^{-n}\binom{n}{a}$. The transverse field mixer has period $\pi$ up to a global phase. Direct integration over this period yields
    \begin{equation} \label{eq:M_TF}
    M_{ab}^{\mathrm{TF}}
    =
    \begin{cases}
    \displaystyle
    (-1)^{(b-a)/2}
    \frac{\binom{n}{a}\binom{n}{b}}{16^n}
    \frac{(a+b)!(2n-a-b)!}
    {
    \left(\frac{a+b}{2}\right)!
    \left(n-\frac{a+b}{2}\right)!
    n!},
    & a+b\ \mathrm{even},\\[12pt]
    0,
    & a+b\ \mathrm{odd}.
    \end{cases}
    \end{equation}

    \item \textbf{Grover mixer.}

    The two mixer classes have mixer factors given by Eq.~\eqref{eq:GM_mixerfactors}, and their marginal probabilities are $P_K^{(z)}(0)=1/N$ and $P_K^{(z)}(1)=(N-1)/N$.  The Grover mixer has period $2\pi$. Integration over this period yields
    \begin{equation} \label{eq:M_GM}
    M^{\mathrm{GM}}
    =
    \frac{1}{N^2}
    \begin{pmatrix}
    \displaystyle 1-\frac{2}{N}+\frac{2}{N^2}
    &
    \displaystyle (N-1)\left(-\frac{1}{N}+\frac{2}{N^2}\right)
    \\[8pt]
    \displaystyle (N-1)\left(-\frac{1}{N}+\frac{2}{N^2}\right)
    &
    \displaystyle \frac{2(N-1)^2}{N^2}
    \end{pmatrix}.
    \end{equation}
\end{itemize}
In particular, for the transverse field mixer, Eq.~(5) of Chernyavskiy et al.~\cite{chernyavskiy2024entropic} provides an equivalent expression for the average probability. Our formulation can be recovered from their result by specializing to a cost value attained by a unique solution and reorganizing the resulting expression in terms of $G_P^{(z)}$ and $M^{\mathrm{TF}}$. 

\subsection{A structural interpretation of QAOA performance}

Our framework can be viewed as providing a complementary perspective on the structural characterization of QAOA developed in Section~\ref{sec:structure}. While the previous characterization is more general and describes the parameter-dependent QAOA amplitude through $P^{(z)}$, the present framework focuses on the setting where polynomial mean--max proximity holds, using $G_P^{(z)}$ to characterize the contribution of problem structure to the asymptotic scaling of the optimal measurement probability, up to polynomial factors. Moreover, as discussed above, the mapping from $P^{(z)}$ to $G_P^{(z)}$ is generally non-injective, reflecting a reduction in the structural information retained by the Gram representation.

Comparing Eqs.~\eqref{eq:amplitude} and~\eqref{eq:overlap} reveals a further distinction concerning the dynamical contributions of the QAOA operators. In the general amplitude representation, both operators enter explicitly through the factors $e^{-i\gamma j}$ and $B_\beta^{(z)}(k)$, respectively. In contrast, the structural overlap contains an explicit dynamical contribution only from the mixer, through $M^{(z)}$. This apparent asymmetry raises a natural question: what happens to the contribution of the problem operator under parameter averaging? To address this question, we return to Eq.~\eqref{eq:amplitude} and replace the problem phase factor $e^{-i\gamma j}$ with a general function $A_\gamma(j)$, recovering standard QAOA when $A_\gamma(j)=e^{-i\gamma j}$.

Let $P^{(z)}$ denote the matrix with entries $[P^{(z)}]_{jk}=P^{(z)}(j,k)$, and define the vectors $[A_\gamma]_j=A_\gamma(j)$ and $[B_\beta^{(z)}]_k=B_\beta^{(z)}(k)$. The amplitude can then be written as $\psi_z(\gamma,\beta)=\sqrt{N}\,A_\gamma^{\mathsf T}P^{(z)}B_\beta^{(z)}$. We introduce the averaged dynamical matrices associated with the two operators, $\Gamma=\mathbb E_\gamma[A_\gamma^* A_\gamma^{\mathsf T}]$ and $M_0^{(z)}=\mathbb E_\beta[B_\beta^{(z)}(B_\beta^{(z)})^\dagger]$. Assuming independent averaging over $\gamma$ and $\beta$, we obtain
\begin{equation}
\overline q_z
=N\operatorname{Tr}\left((P^{(z)})^{\mathsf T}\Gamma P^{(z)}M_0^{(z)}\right).
\end{equation}
Thus, in this more general representation, both operators contribute explicitly through their respective averaged dynamical matrices.

For QAOA, the problem factor is $A_\gamma(j)=e^{-i\gamma j}$. Since the indices $j$ label distinct problem costs, long-time averaging gives
\begin{equation}
(\Gamma)_{jj'}
=
\mathbb E_\gamma[e^{i\gamma(j-j')}]
=
\delta_{jj'},
\qquad\text{hence}\qquad \Gamma=I.
\end{equation}
Consequently, $\overline q_z=N\operatorname{Tr}\left((P^{(z)})^{\mathsf T}P^{(z)}M_0^{(z)}\right)$. This expression is equivalent to Eq.~\eqref{eq:overlap}, since $P^{(z)}=P^{(z)}_{J\mid K}P_K^{(z)}$ and therefore $(P^{(z)})^{\mathsf T}P^{(z)}=P_K^{(z)}G_P^{(z)}P_K^{(z)}$.

The apparent asymmetry is therefore a consequence of the phase structure of the standard QAOA problem operator. Averaging over $\gamma$ eliminates interference between distinct cost classes while preserving contributions associated with equal costs, reducing its averaged dynamical matrix to the identity. However, the problem contribution is not lost: the distribution of equal-cost classes across mixer classes remains encoded in $G_P^{(z)}$, while the mixer retains nontrivial averaged dynamical correlations through $M^{(z)}$.

Therefore, we can conclude that the asymptotic scaling of the optimal measurement probability is determined, up to polynomial factors, by the interaction between the relevant problem structure, encoded in $G_P^{(z)}$, and the intrinsic properties of the mixer, captured by $M^{(z)}$.

\subsection{Structural competition and superpolynomial suppression}

We now examine the structural overlap by classifying the individual terms in its defining sum. We then investigate their asymptotic behavior, leading to the identification of a general mechanism of superpolynomial suppression in QAOA.

\subsubsection{Structural classification}

Recall that the structural overlap is given by
\begin{equation}
\label{eq:overlap-decomposition}
\langle G_P^{(z)},M^{(z)}\rangle_{\mathrm{HS}}
=
\sum_{a,b}(G_P^{(z)})_{ab}M_{ab}^{(z)}.
\end{equation}
Since both matrices are real and $(G_P^{(z)})_{ab}\geq 0$, the sign of each contribution is determined by $M_{ab}^{(z)}$. Thus, based on the entries of $G_P^{(z)}$, we distinguish three types of problem structure:
\begin{itemize}
    \item \textbf{Constructive structure:} $M_{ab}^{(z)}>0$, contributing positively to the structural overlap.

    \item \textbf{Blind structure:} $M_{ab}^{(z)}=0$, contributing nothing to the structural overlap.

    \item \textbf{Destructive structure:} $M_{ab}^{(z)}<0$, contributing negatively to the structural overlap.
\end{itemize}
A problem may exhibit a mixed structure, with constructive, blind, and destructive entries of $G_P^{(z)}$, whose classification depends on the chosen mixer. For the transverse field mixer, $M_{ab}^{(z)}=0$ when $a+b$ is odd, while for $a+b$ even its sign is determined by $(-1)^{(b-a)/2}$. Thus, constructive and destructive contributions alternate according to the Hamming-distance classes, with blind regions separating them. For the Grover mixer, the diagonal entries of $M^{(z)}$ are positive, while the off-diagonal entries are negative (for $N>2$). Consequently, the diagonal structure is constructive and the off-diagonal structure is destructive, with no blind regions.

\subsubsection{Asymptotic competition between constructive and destructive contributions}

Under the polynomial complexity conditions $m,d,\ell=\operatorname{poly}(n)$, the structural overlap determines the optimal probability up to polynomial factors and contains only polynomially many terms. Consequently, superpolynomial rates are the relevant quantities for our asymptotic analysis, with the largest constructive and destructive terms playing a central role. To make this explicit, we decompose the structural overlap into two nonnegative contributions:
\begin{equation}
\langle G_P^{(z)},M^{(z)}\rangle_{\mathrm{HS}}
=
\mathcal{C}-\mathcal{D},
\end{equation}
where
\begin{equation}
\mathcal{C}
=
\sum_{M_{ab}^{(z)}>0}(G_P^{(z)})_{ab}M_{ab}^{(z)},
\qquad
\mathcal{D}
=
\sum_{M_{ab}^{(z)}<0}(G_P^{(z)})_{ab}|M_{ab}^{(z)}|.
\end{equation}
Let
\begin{equation}
C_{\max}
=
\max_{M_{ab}^{(z)}>0}(G_P^{(z)})_{ab}M_{ab}^{(z)},
\qquad
D_{\max}
=
\max_{M_{ab}^{(z)}<0}(G_P^{(z)})_{ab}|M_{ab}^{(z)}|.
\end{equation}
Since the number of terms is polynomially bounded, we have
\begin{equation}
\mathcal{C}\overset{\mathrm{poly}}{\asymp}C_{\max},
\qquad
\mathcal{D}\overset{\mathrm{poly}}{\asymp}D_{\max},
\end{equation}
whenever the corresponding contributions are nonzero.

Recall that the structural overlap is always positive, so $\mathcal C > \mathcal D$. Based on the relative scaling of these contributions, we distinguish three asymptotic regimes of particular interest.
\begin{definition}[Asymptotic regimes of structural competition]
\label{def:asymptotic}
Consider the decomposition $\langle G_P^{(z)},M^{(z)}\rangle_{\mathrm{HS}}=\mathcal C-\mathcal D$. We distinguish the following regimes:
\begin{enumerate}
    \item \textit{Constructive dominance:} $\mathcal D\mathrel{\not\asymp^{\mathrm{poly}}}\mathcal C$.

    \item \textit{Balanced competition without superpolynomial suppression:} $\mathcal D\overset{\mathrm{poly}}{\asymp}\mathcal C$ and $\mathcal C-\mathcal D\overset{\mathrm{poly}}{\asymp}\mathcal C$.

    \item \textit{Balanced competition with superpolynomial suppression:} $\mathcal D\overset{\mathrm{poly}}{\asymp}\mathcal C$ and $(\mathcal C-\mathcal D)/\mathcal C=n^{-\omega(1)}$.
\end{enumerate}
\end{definition}
These regimes characterize the main asymptotic patterns of structural competition. They need not exhaust all possibilities, since asymptotically irregular families may exhibit scaling behavior that falls outside the polynomial and superpolynomial classifications considered here.

Note that blind regions do not contribute directly to this competition. Nevertheless, they may contain significant entries of $G_P^{(z)}$ that could become dynamically accessible under a different mixer.

\subsubsection{The mechanism of superpolynomial suppression}

We observe that the matrix $M^{(z)}$ possesses an intrinsic capacity that bounds the structural overlap independently of the objective function. Indeed, the triangle inequality and the elementwise bounds on $G_P^{(z)}$ imply
\begin{equation}
\label{eq:mixer-overlap-bound}
\langle G_P^{(z)},M^{(z)}\rangle_{\mathrm{HS}}
\leq
\|M^{(z)}\|_1,
\end{equation}

Since this quantity depends only on the mixer and the reference vertex, we refer to it as the \emph{mixer capacity}. For the transverse field and Grover mixers, Eqs.~\eqref{eq:M_TF} and \eqref{eq:M_GM} yield the asymptotic scaling of the corresponding mixer capacities:
\begin{equation}
\|M^{\mathrm{TF}}\|_1\overset{\mathrm{poly}}{\asymp} \frac{1}{N},
\qquad
\|M^{\mathrm{GM}}\|_1\overset{\mathrm{poly}}{\asymp}\frac{1}{N^{2}}.
\end{equation}
For the transverse field mixer, the scaling follows by considering the diagonal entry $a=b=\lfloor n/2\rfloor$, while for the Grover mixer, the scaling follows directly from Eq.~\eqref{eq:M_GM}. Comparing these results with the universal bounds in Eq.~\eqref{eq:universal-overlap-bounds}, we observe that the transverse field mixer has sufficient capacity to reach the maximal scale of the structural overlap, up to polynomial factors. An explicit example is the Hamming weight problem $c(x)=|x|$, with reference state $z=0^n$, for which one can directly verify that $\psi_z(\pi/2,-\pi/4)=1$. In contrast, the Grover mixer is intrinsically restricted to the minimal scale allowed by the universal bounds, up to polynomial factors, consistent with previous quadratic speed-up bounds~\cite{bridi2024analytical,xie2025performance}.

The following theorem establishes that the constructive contribution saturates the mixer capacity up to polynomial factors. Consequently, any additional superpolynomial suppression of the structural overlap must arise from cancellations between constructive and destructive contributions.

\begin{theorem}[Constructive saturation of mixer capacity]
\label{thm:constructive-saturation}
Consider single-layer QAOA initialized in the uniform superposition $|s\rangle$, with a real-valued cost function $c$ taking $m$ distinct values and an arbitrary Hermitian mixer Hamiltonian $H_M$ with $d$ distinct eigenvalues and inducing $\ell$ mixer classes. For any reference configuration $z$, the constructive contribution satisfies
\begin{equation}
\label{eq:constructive-saturation}
\frac{\|M^{(z)}\|_1}{m\ell}
\leq
\mathcal C
\leq
\|M^{(z)}\|_1.
\end{equation}
In particular, for a family of problem instances associated with an NPO-PB problem of size $n$, using a mixer Hamiltonian with a polynomially bounded number of distinct eigenvalues, $d=\operatorname{poly}(n)$, and a polynomially bounded number of mixer classes, $\ell=\operatorname{poly}(n)$, we have
\begin{equation}
\mathcal C
\overset{\mathrm{poly}}{\asymp}
\|M^{(z)}\|_1.
\end{equation}
\end{theorem}

\begin{proof}
Since $M_{aa}^{(z)}\geq 0$, the diagonal contribution provides a lower bound for the constructive contribution $\mathcal C$ as $\mathcal C \geq \sum_a (G_P^{(z)})_{aa}M_{aa}^{(z)}$. By the definition of $G_P^{(z)}$, $(G_P^{(z)})_{aa}= \sum_j P^{(z)}_{J|K}(j|a)^2$. The Cauchy--Schwarz inequality gives $(G_P^{(z)})_{aa}\geq\frac{1}{m}$. Consequently,
\begin{equation}
\label{eq:constructive-diagonal-bound}
\mathcal C
\geq
\frac{\operatorname{Tr}(M^{(z)})}{m}.
\end{equation}
Furthermore, positive semidefiniteness of $M^{(z)}$ implies $|M_{ab}^{(z)}|\leq\sqrt{M_{aa}^{(z)}M_{bb}^{(z)}}\leq\frac{M_{aa}^{(z)}+M_{bb}^{(z)}}{2}$. Summing over all pairs $(a,b)$ yields
\begin{equation}
\|M^{(z)}\|_1
\leq
\ell\,\operatorname{Tr}(M^{(z)}).
\end{equation}
Combining these inequalities establishes the lower bound in Eq.~\eqref{eq:constructive-saturation}. The upper bound follows directly from $(G_P^{(z)})_{ab}\leq 1$ and the definition of $\mathcal C$.
\end{proof}

Under the polynomial boundedness conditions of Theorem~\ref{thm:constructive-saturation}, superpolynomial suppression of the structural overlap relative to the mixer capacity can only arise from cancellations between constructive and destructive contributions. Consequently, in the constructive dominance regime, the structural overlap retains the superpolynomial scaling classification of the mixer capacity, as formalized in the following corollary.

\begin{corollary}[Constructive dominance and mixer capacity]
\label{cor:constructive-dominance}
Consider single-layer QAOA initialized in the uniform superposition $|s\rangle$, for a family of problem instances associated with an NPO-PB problem of size $n$, using an arbitrary Hermitian mixer Hamiltonian $H_M$ with a polynomially bounded number of distinct eigenvalues, $d=\operatorname{poly}(n)$, and a polynomially bounded number of mixer classes, $\ell=\operatorname{poly}(n)$. For any reference configuration $z$, if the structural overlap is in the constructive dominance regime (Definition~\ref{def:asymptotic}), then
\begin{equation}
\label{eq:constructive-dominance-capacity}
\langle G_P^{(z)},M^{(z)}\rangle_{\mathrm{HS}}
\overset{\mathrm{poly}}{\asymp}
\|M^{(z)}\|_1.
\end{equation}
\end{corollary}

\subsubsection{Canonical examples of competition}

We now illustrate the three asymptotic regimes of structural competition through explicit problem--mixer pairs, examining how the interplay between constructive and destructive contributions determines the superpolynomial scaling classification of the structural overlap. In these examples, we focus specifically on exponential scaling.

\paragraph{Constructive dominance.}

Consider first a cost function whose classes are an injective function of the mixer classes, such that $J=f(K)$ for some injective map $f$. By the definition of $G_P^{(z)}$ and the injective invariance of $G_P^{(z)}$ established in Eq.~\eqref{eq:GP-injective-invariance}, we obtain $G_P^{(z)}=I$. Consequently, the structural overlap receives contributions only from the diagonal entries of $M^{(z)}$, which are necessarily nonnegative and therefore contribute constructively. By Corollary~\ref{cor:constructive-dominance}, the structural overlap satisfies $\langle G_P^{(z)},M^{(z)}\rangle_{\mathrm{HS}}\overset{\mathrm{poly}}{\asymp}\|M^{(z)}\|_1$. This case includes the transverse field mixer applied to the Hamming weight problem $c(x)=|x|$, with reference state $z=0^n$, and the Grover mixer applied to unstructured search, with $z$ being the marked element.

A concrete example satisfying the hypotheses of Corollary~\ref{cor:constructive-dominance} in a more general setting, beyond the diagonal case, is provided by the transverse field mixer with the cost function $J=K\bmod 4$, where $K$ labels the Hamming-distance classes relative to $z=0^n$. The corresponding problem structure matrix is
\begin{equation}
(G_P^{(z)})_{ab}
=
\begin{cases}
1, & a\equiv b\pmod 4,\\
0, & \text{otherwise}.
\end{cases}
\end{equation}
From Eq.~\eqref{eq:M_TF}, the nonzero entries of $M^{\mathrm{TF}}$ have signs determined by $(-1)^{(b-a)/2}$. Since $a\equiv b\pmod 4$ implies $(-1)^{(b-a)/2}=1$, the problem structure selects only nonnegative contributions. 

\paragraph{Balanced competition without superpolynomial suppression.}

Consider the Max-Cut problem on the complete bipartite graph $K_{L,L}$, with $n=2L$ and $L$ odd, using the transverse field mixer. The Max-Cut objective counts the number of edges connecting vertices assigned different binary values. Since the graph is bipartite, flipping the bits of one partition maps an optimal cut to $z=0^n$ without changing the corresponding QAOA measurement probabilities, provided the cost function is transformed accordingly. We may therefore work, without loss of generality, in this equivalent representation, where the cost function is
\begin{equation}
c(x)=L^2-L(u+v)+2uv,
\qquad
u=|x_A|,
\quad
v=|x_B|.
\end{equation}
Here, $x_A$ and $x_B$ denote the bit strings associated with the two partitions, and $c(0^n)=L^2$ is the maximum cut value.

We first establish the exponential scale of the structural overlap. Choosing $\gamma=\pi/2$ and $\beta=\pi/4$ for $L\equiv1\pmod4$, or $\beta=-\pi/4$ for $L\equiv3\pmod4$ and substituting these parameters into the reference state amplitude in Eq.~\eqref{eq:unif_amplitude}, we have
\begin{equation}
q_z
=
\frac{1}{N^2}
\left|
\sum_{u,v=0}^{L}
\binom Lu\binom Lv(-1)^{uv}
\right|^2.
\end{equation}
Separating the sum into even and odd values of $u$, and using $\sum_v\binom Lv(-1)^v=0$ and $\sum_{u\text{ even}}\binom Lu=2^{L-1}$, we obtain $q_z=1/4$. Since Max-Cut is an NPO-PB problem and the transverse field mixer has a polynomially bounded number of distinct eigenvalues, Theorem~\ref{thm:structural-average} implies
$\langle G_P^{(z)},M^{\mathrm{TF}}\rangle_{\mathrm{HS}}
\overset{\mathrm{poly}}{\asymp} N^{-1}$,
which is the same exponential scale as $\|M^{\mathrm{TF}}\|_1$.

We now show that the destructive contribution has the same exponential scale. Consider the mixer classes $a=L-1$ and $b=L+1$. From Eq.~\eqref{eq:M_TF}, $(G_P^{(z)})_{L-1,L+1}$ is destructive structure. Moreover, configurations with weights $(u,v)$ satisfying $u+v=L-1$ have the same cost as configurations with weights $(u+1,v+1)$. In particular, taking $u=v=(L-1)/2$, the corresponding conditional probabilities are only polynomially small, implying
\begin{equation}
(G_P^{(z)})_{L-1,L+1}
\geq
\left[
\frac{\binom{L}{(L-1)/2}^{2}}
{\binom{2L}{L-1}}
\right]^2
=
\Omega\left(\frac{1}{L}\right).
\end{equation}
Furthermore, the explicit expression for $M^{\mathrm{TF}}$ gives
\begin{equation}
|M^{\mathrm{TF}}_{L-1,L+1}|
=
\frac{\binom{2L}{L-1}^{2} \binom{2L}{L}}{N^4}
=
\Theta\left(\frac{1}{NL^{3/2}}\right).
\end{equation}
Consequently, this single negative contribution provides the lower bound
\begin{equation}
\mathcal D
\geq
(G_P^{(z)})_{L-1,L+1}|M^{\mathrm{TF}}_{L-1,L+1}|
=
\Omega\left(\frac{1}{NL^{5/2}}\right).
\end{equation}
Since $\mathcal D\leq\|M^{\mathrm{TF}}\|_1\overset{\mathrm{poly}}{\asymp}N^{-1}$, it follows that $\mathcal D\overset{\mathrm{poly}}{\asymp}N^{-1}$.

Combining these results with Theorem~\ref{thm:constructive-saturation}, which guarantees that $\mathcal C\overset{\mathrm{poly}}{\asymp}\|M^{\mathrm{TF}}\|_1\overset{\mathrm{poly}}{\asymp}N^{-1}$, we conclude that the Max-Cut problem on $K_{L,L}$ with odd $L$ provides an explicit example of balanced competition without superpolynomial suppression. In particular, constructive and destructive contributions coexist at the same scale, but their cancellation does not produce any additional superpolynomial suppression of the structural overlap.

\paragraph{Balanced competition with superpolynomial suppression.}

Finally, consider the transverse field mixer applied to the unstructured search problem, with $J(x)=\delta_{x,z}$ and $z=0^n$. In this case, the problem structure matrix takes the form
\begin{equation}
(G_P^{(z)})_{ab}
=
\begin{cases}
1, & a=b=0,\\
1, & a,b\geq 1,\\
0, & \text{otherwise}.
\end{cases}
\end{equation}

Notice that part of the expression for $M^{\mathrm{TF}}_{ab}$ in Eq.~\eqref{eq:M_TF} depends only on the sum $a+b$. This suggests grouping the terms according to their common value of $a+b$, which corresponds to grouping the matrix entries along antidiagonals. Setting $a+b=2r$, we can write
\begin{equation}
M^{\mathrm{TF}}_{ab}
=
h_r(-1)^{(b-a)/2}\binom na\binom nb,
\qquad
h_r=
\frac{(2r)!(2n-2r)!}
{16^n r!(n-r)!n!}.
\end{equation}
Consequently, the structural overlap can be expressed as a sum over antidiagonals,
\begin{equation}
\langle G_P^{(z)},M^{\mathrm{TF}}\rangle_{\mathrm{HS}}
=
\sum_{r=0}^{n}h_r A_r, \quad A_r =
\sum_{a+b=2r}
(G_P^{(z)})_{ab}(-1)^{(b-a)/2}
\binom na\binom nb.
\end{equation}
Observe that $h_r$ is strictly positive, whereas $A_r$ contains the alternating sign $(-1)^{(b-a)/2}$, which allows for cancellations between constructive and destructive contributions within each antidiagonal. For unstructured search, the structure of $G_P^{(z)}$ allows us to evaluate $A_r$ explicitly, yielding
\begin{equation}
\begin{aligned}
A_r
&=
\sum_{a+b=2r}
(-1)^{(b-a)/2}\binom na\binom nb
-2(-1)^r\binom n{2r}
+2\delta_{r0}\\
&=
(-1)^r[x^{2r}](1-x)^n(1+x)^n
-2(-1)^r\binom n{2r}
+2\delta_{r0}\\
&=
\binom nr-2(-1)^r\binom n{2r}+2\delta_{r0},
\end{aligned}
\end{equation}
where $\binom n{2r}=0$ whenever $2r>n$. Consequently, the structural overlap becomes
\begin{equation}
\langle G_P^{(z)},M^{\mathrm{TF}}\rangle_{\mathrm{HS}}
=
\sum_{r=0}^{n}
h_r
\left[
\binom nr
-2(-1)^r\binom n{2r}
+2\delta_{r0}
\right].
\end{equation}
Using the identities
\begin{equation}
\sum_{r=0}^{n}h_r\binom nr=\frac1{N^2},
\qquad
\sum_{r=0}^{n}(-1)^rh_r\binom n{2r}=\frac1{N^3},
\end{equation}
which follow from standard binomial identities and coefficient extraction, together with $h_0=\binom{2n}{n}/N^4$, we obtain
\begin{equation}
\langle G_P^{(z)},M^{\mathrm{TF}}\rangle_{\mathrm{HS}}
=
\frac1{N^2}
\left[
1-\frac2N+
2\frac{\binom{2n}{n}}{N^2}
\right]
\overset{\mathrm{poly}}{\asymp} \frac{1}{N^{2}}.
\end{equation}
The alternating signs of the transverse field mixer produce cancellations within the antidiagonal sums, suppressing the structural overlap to the scale $N^{-2}$ despite the mixer capacity $\|M^{\mathrm{TF}}\|_1\overset{\mathrm{poly}}{\asymp}N^{-1}$. By Theorem~\ref{thm:constructive-saturation}, $\mathcal C\overset{\mathrm{poly}}{\asymp}N^{-1}$, and therefore $\mathcal D\overset{\mathrm{poly}}{\asymp}N^{-1}$, establishing balanced competition with superpolynomial suppression.

\section{Conclusions}

The central result of this work is a general mean--max bound that quantifies how much variational optimization can improve QAOA measurement probabilities for arbitrary subsets of computational basis configurations. Under polynomial mean--max proximity, this bound has two major implications, one algorithmic and the other structural.

From an algorithmic perspective, our results clarify the role of variational optimization in QAOA. Under polynomial mean--max proximity and suitable efficiency assumptions, the ability to sample desirable solutions does not fundamentally depend on identifying optimized parameters. From a structural perspective, polynomial mean--max proximity constrains the impact of variational optimization to polynomial factors, revealing a separation between the structural factors governing the asymptotic exponential scale of measurement probabilities and the improvements achievable through parameter optimization. For single-layer QAOA initialized in the uniform superposition state, we make this explicit by identifying a pre-optimization structural layer that determines the asymptotic scaling of the optimal measurement probability of a reference state up to polynomial factors. This characterization further reveals how problem structure and mixer dynamics jointly shape QAOA performance.

\vspace{0.4cm}
\paragraph{Acknowledgments.} G.A.B. acknowledges support from CAPES, Finance Code 001. F.L.M. acknowledges support from CNPq Productivity Grant. G.A.B. and F.L.M. acknowledge support of the National Institute of Science and Technology for Applied Quantum Computing. A.A. acknowledges the QuantERA project HQCC (Hybrid Quantum Classical Computation). The authors used ChatGPT to check grammar and to improve the clarity and organization of the text. The authors reviewed all suggestions and take full responsibility for the content of the manuscript.

\bibliographystyle{unsrt}
\bibliography{references}{}

\end{document}